\documentclass[reprint,superscriptaddress,amsmath,amssymb,amsfonts,pra]{revtex4-2}
\usepackage{graphicx}% Include figure files
\usepackage{dcolumn}% Align table columns on decimal point
\usepackage{bm}% bold math
\usepackage{mathrsfs}
\usepackage{xcolor}
\usepackage{physics}
\usepackage{xr-hyper}

\usepackage{rotating} % \begin{turn}{90}; \rotatebox{90}{...}
\usepackage{varwidth} % warp up the table and caption 
\usepackage{threeparttable} % table footnote
\usepackage{subfigure}

\usepackage[normalem]{ulem} % striking out the text
\usepackage{soul} % striking out the text  % obsolete!

\usepackage[colorlinks=true]{hyperref}

\newcommand{\response}[1]{\textcolor{black}{#1}} % blue % Response to Referee

\newcommand{\pdfproof}[1]{\textcolor{black}{#1}} % red % Response to Editor

\makeatletter
\def\maketitle{
\@author@finish
\title@column\titleblock@produce
\suppressfloats[t]}
\makeatother

\begin{document}

% Title
%\title{Quantum Computing with DFT Molecule Orbital expanded in the Daubechies Wavelet Basis Set}
\title{\pdfproof{Accurate harmonic vibrational frequencies for diatomic molecules via \\ quantum computing}}

\author{Shih-Kai Chou}
\affiliation{Department of Physics and Center for Theoretical Physics, National Taiwan University, Taipei 10617, Taiwan}

\author{Jyh-Pin Chou}
\affiliation{Department of Physics, National Changhua University of Education, Changhua 50007, Taiwan}
\affiliation{Physics Division, National Center for Theoretical Sciences, Taipei, 10617, Taiwan}

\author{Alice Hu}
\affiliation{Department of Mechanical Engineering, City University of Hong Kong, Kowloon, Hong Kong SAR 999077, China}
\affiliation{Department of Materials Science and Engineering, City University of Hong Kong, Kowloon, Hong Kong SAR 999077, China}

\author{Yuan-Chung Cheng}
\email{yuanchung@ntu.edu.tw}
\affiliation{Department of Chemistry, National Taiwan University, Taipei 10617, Taiwan}
\affiliation{Center for Quantum Science and Engineering, National Taiwan University, Taipei 10617, Taiwan}
\affiliation{Physics Division, National Center for Theoretical Sciences, Taipei, 10617, Taiwan}

\author{Hsi-Sheng Goan}
\email{goan@phys.ntu.edu.tw}
\affiliation{Department of Physics and Center for Theoretical Physics, National Taiwan University, Taipei 10617, Taiwan}
\affiliation{Center for Quantum Science and Engineering, National Taiwan University, Taipei 10617, Taiwan}
\affiliation{Physics Division, National Center for Theoretical Sciences, Taipei, 10617, Taiwan}

%\date{\today}

%------------------------------------------------------------------------------
%  Abstruct
%------------------------------------------------------------------------------
\begin{abstract}
During the noisy intermediate-scale quantum (NISQ) era, quantum computational approaches refined to overcome the challenge of limited quantum resources are highly valuable. A comprehensive benchmark for a quantum computational approach in this spirit could provide insights toward further improvements. On the other hand, the accuracy of the molecular properties predicted by most of the quantum computations nowadays is still far off (not within chemical accuracy) compared to their corresponding experimental data. In this work, we propose a promising qubit-efficient quantum computational approach and present a comprehensive investigation by benchmarking quantum computation of the harmonic vibrational frequencies of a large set of neutral closed-shell diatomic molecules with results in great agreement with their experimental data. To this end, we construct the accurate Hamiltonian using molecular orbitals, derived from density functional theory to account for the electron correlation and expanded in the Daubechies wavelet basis set to allow an accurate representation in real space grid points, where an optimized compact active space is further selected so that only a reduced small number of qubits is sufficient to yield an accurate result. Typically, calculations achieved with 2 to 12 qubits using our approach would need 20 to 60 qubits using \pdfproof{a} traditional cc-pVDZ basis set with frozen core approximation to achieve similar accuracy. To justify the approach, we benchmark the performance of the Hamiltonians spanned by the selected molecular orbitals by first transforming the molecular Hamiltonians into qubit Hamiltonians and then using the exact diagonalization method to calculate the results, regarded as the best results achievable by quantum computation to compare to the experimental data. Furthermore, using the variational quantum eigensolver algorithm with the constructed qubit Hamiltonians, we show that the variational quantum circuit with the chemistry-inspired UCCSD ansatz can achieve the same accuracy as the exact diagonalization method except for systems whose Mayer bond order indices are larger than 2. For those systems, we then demonstrate that the heuristic hardware-efficient RealAmplitudes ansatz, even with a substantially shorter circuit depth, can provide \pdfproof{a} significant improvement over the UCCSD ansatz, verifying that the harmonic vibrational frequencies could be calculated accurately by quantum computation in the NISQ era.
\end{abstract}

\maketitle

%------------------------------------------------------------------------------
%  INTRODUCTION
%------------------------------------------------------------------------------
\section{INTRODUCTION}
Recently, quantum computing has emerged as a promising way to potentially solve classically intractable problems, especially in the field of quantum chemistry, whose fundamental goal is to solve the Schr\"{o}dinger equation for chemical or molecular systems \cite{Cao:CR2019,McArdle:RMP2020,Bharti:RMP2022}. With the quantum nature of \pdfproof{wave functions}, quantum computing makes use of superposition, entanglement and interference to prepare and manipulate quantum states, offering the potential for exponential speedup over classical computing. Along with this quantum advantage, the unitary coupled cluster with single and double excitations (UCCSD) method \cite{UCC:Kutzelnigg1977, UCC:Hoffmann1988, UCC:Bartlett1989} for quantum computation of the molecular properties can be realized by mapping the exponential operators to qubit operators and using \pdfproof{t}rotterization to approximate the corresponding quantum circuit. This circuit can be efficiently implemented on a quantum computer \cite{Peruzzo:NC2014,Yung:SE2014,Romero:QST2018,Anand:CSR2022}, making UCCSD a powerful tool for simulating chemical systems. The number of gates for the UCCSD circuit, using the Jordan-Wigner transformation \cite{JW}, scales as $\mathcal{O}(N_q^3 N_e^2)$, where $N_q$ is the number of qubits and $N_e$ is the number of electrons \cite{Romero:QST2018}. On the other hand, classical implementation of the UCCSD method is impractical due to the nontruncated Baker-Campbell-Hausdorff expansion \cite{Taube:IJQC2006}. In addition, the gold standard method in quantum chemistry by classical computation, the coupled cluster with single, double, and perturbative triple excitations $($CCSD(T) \cite{CCSDt}$)$, scaling as $\mathcal{O}(N^7)$, where $N$ is the number of molecular orbitals (MOs), is applicable only to small systems. Therefore the UCCSD method is considered a promising candidate for the quantum simulation, and its efficient implementation is a major area of research in the field of quantum computational chemistry. As quantum computing technology continues to advance, it is expected that the UCCSD method will play an increasingly important role in the development of applications of quantum chemistry.

In the so-called noisy intermediate-scale quantum (NISQ) era \cite{Preskill:NISQ}, quantum computers have noisy and a limited number of qubits without error correction implemented. As a result, the number of consecutive quantum gates that can be reliably run on the NISQ machines is also restricted. To address this challenge, a hybrid quantum-classical algorithm called the variational quantum eigensolver (VQE) \cite{Peruzzo:NC2014} has been proposed and widely used. 
In this VQE framework, a parameterized quantum circuit as an ansatz to prepare the trial quantum states (\pdfproof{wave functions}) is optimized to find the ground state energy of the Hamiltonian based on the variational principle.
The VQE algorithm leverages the strengths of quantum and classical computation, and distributes the computational workload between quantum and classical computers, with the quantum computer performing the trial quantum state  preparation and measurement, and the classical computer performing the parameter optimization.
 The structure of the circuit ansatz and the ability of parameter optimization determine the accuracy of the result obtained from the VQE algorithm. 

Reducing the requirement on the number of qubits and thus the depth of the quantum circuit is one of the major strategies in the NISQ era. 
For chemical problems, the chemistry-inspired UCCSD ansatz is commonly used. However, due to its complex circuit structure resulting from an exponential operator, the UCCSD ansatz can give rise to a very deep quantum circuit, making it difficult to be really implemented on NISQ devices. In contrast,  the heuristic hardware-efficient ansatz is proposed to take advantage of its shorter circuit depth than that of the chemistry-inspired ansatz on NISQ devices \cite{Kandala:Nature2017}. The general construction of a hardware-efficient ansatz consists of alternating layers of parameterized single-qubit rotation gates and two-qubit entangling gates. 
In general, the true \pdfproof{wave function} of a quantum system can be expressed as a unitary transformation of the initial state \cite{Taube:IJQC2006}. Despite it is not guaranteed that a hardware-efficient ansatz of a unitary quantum circuit contains the solution of the \pdfproof{wave function} or it is optimal, and that it preserves the same properties as the true \pdfproof{wave function} \cite{Barkoutsos:PRA2018, Gard:npjQI2020}, a hardware-efficient ansatz is flexible to vary the types and increase the number of layers of parameterized and entangling gates, allowing it to cover more of the solution space where the true \pdfproof{wave function} may reside. Some reviews about the recent development for the both kinds of ansatzes could be found in \cite{Cerezo:NRP2021, Fedorov:MT2022}.
%However, the current implementation of a hardware-efficient ansatz does not guarantee that it contains the true solution or that it is optimal. To ensure that a hardware-efficient ansatz preserves the same properties as the true wave function, the specific architecture should be incorporated \cite{Barkoutsos:PRA2018, Gard:npjQI2020}. Some reviews about recent development for the both types of ansatzes could be found in \cite{Cerezo:NRP2021, Fedorov:MT2022}.
Recently, an adaptive variational algorithm, called adaptive derivative-assembled pseudo-Trotter ansatz VQE (ADAPT-VQE) \cite{Grimsley:NatCommun2019}, has been proposed to 
determine adaptively a \pdfproof{quasioptimal} ansatz with the minimal number of
excitation operators  (e.g. as in the UCCSD ansatz) for molecular simulations.  
Although ADAPT-VQE can yield an adaptive ansatz with a considerably reduced number of parameters, the ansatz circuit that becomes shallower than the UCCSD ansatz is still too deep to be implemented successfully on the current NISQ devices.

Besides the variational ansatz encoded in the trial \pdfproof{wave function}, quantum computation of quantum chemistry also depends on a representation of the molecular Hamiltonian. 
%High-quality representation of the molecular Hamiltonian is essential for accurate predictions of chemical properties, which was often overlooked in the quantum computing community. Generally speaking, a larger basis set makes better representation of the Hamiltonian, but it accompanies the burden of higher computational cost. In order to represent the Hamiltonian in a smaller basis set for the NISQ devices without losing much of the accuracy, the choice of the basis set is thus extremely important. Traditional basis set to express the MOs is constructed from linear combination of atomic orbitals (LCAO) represented by atom-centered basis functions (known as the LCAO-MO approach). Another type of basis set is from the real-space numerical grid method, where the MO is expanded in the set of specific real-space basis functions. The key feature of this method is that the accuracy of the real-space basis set could be systematically improvable as the number of basis functions is increased. Such real-space basis set fulfills the need here and has existed on applications with outstanding performance in the literature.
A common choice as a first demonstration is the minimal basis (MB) set like STO-3G \cite{STO-3G}, but the calculation using the STO-3G MB set usually does not yield an accurate result.
To improve the predicted results to be comparable with their experimental data, a larger basis set such as cc-pVDZ \cite{Dunning_I:JCP1989,Dunning_III:JCP1993,Dunning_IX:JCP1999,Dunning_VII:TCA2011} is widely adopted in classical computation  of quantum chemistry.  However, even for the simplest H$_2$ molecule, the VQE calculation with Hamiltonian in the cc-pVDZ basis set required 20 qubits and the circuit depth estimated by using the UCCSD ansatz would be over $10^4$. Such a very deep circuit for H$_2$ is obviously not realizable on the current NISQ devices, not to mention for larger molecules. In order to represent a high-quality Hamiltonian in a smaller basis set for the NISQ devices without losing much of the accuracy, the choice of the basis set is thus extremely important, which was often overlooked in the quantum computing community.
Different from traditional basis sets, the basis set constructed from the real-space numerical grid method, where the MO is expanded in the set of specific real-space basis functions, could fulfill the need here. 
For example, 
%for the classical computation, the Lagrange-sinc function has been used on the configuration interaction (CI) \cite{CI} calculation by Kim et al \cite{Kim:JCP2015, Kim:PCCP2015, Kim:JCP2016, Kim:JCP2020}. For quantum computation,
the multiresolution analysis (MRA) \cite{Harrison:JCP2004, MADNESS, Kottmann:JCP2020} method has been used to represent pair-nature orbitals as an efficient basis-set-free approach to simulate molecular systems for VQE by Kottmann \pdfproof{\textit{et al.}~}\cite{Kottmann:JPCL2021}.
Although the qubit requirements can be significantly reduced, the MRA approach is treated as a black box, and for the BeH$_2$ molecule, their approach would not yield a smooth potential energy curve (PEC).
More recently, Hong \pdfproof{\textit{et al.}~}\cite{Hong:PRXQ2022} demonstrated that a MB set constructed from Daubechies wavelet \cite{Daubechies} MOs calculated from \pdfproof{\textsc{bigdft}} \cite{Genovese:JCP2008,Genovese:CRM2011,Mohr:JCP2014,Ratcliff:JCP2020} can yield accurate results in harmonic vibrational frequencies for H$_2$, LiH, and BeH$_2$ on quantum simulator with noisy model implemented from the real devices.
%the Daubechies wavelets \cite{Daubechies} has been used by Hong et al. \cite{Hong:PRXQ2022} to demonstrate that harmonic vibrational frequencies of H$_2$, LiH, and BeH$_2$ can be calculated accurately \st{on NISQ devices} \textcolor{orange}{on quantum simulator with noisy model implemented from real devices}.
The key feature of real-space  basis set of, e.g., Daubechies wavelet MOs is that its accuracy could be systematically improvable as the number of basis functions is increased.
It was demonstrated in \pdfproof{Ref.~}\cite{Hong:PRXQ2022} that VQE quantum computations for vibrational frequencies using the MB set of Daubechies wavelet MOs with accuracy comparable with that of the full configuration interaction (FCI) \cite{CI} calculation using the cc-pVDZ basis set, whereas the computational cost the same as that of a STO-3G calculation, have been achieved for this small set of three simple molecules. 
However, while further inspecting on the application of Daubechies wavelet basis set extended to a larger benchmark molecular dataset considered here (see \pdfproof{Sec.~}\ref{sec:dataset}), we find that the approach of using the MB set of Daubechies wavelet MOs \cite{Hong:PRXQ2022} does not provide adequate results of vibrational frequencies as it fails to produce smooth PECs for some molecules, and furthermore by excluding those unavailable vibrational frequency data, it also has a significantly larger root-mean-square deviation (RMSD) value even though the number of active MOs used is considerably higher than that of our proposed approach here (see \pdfproof{Ref.~}\cite{SupplementalMaterial}). This intrigues a study of active space since important orbitals may not be inside the MB set.

Selecting active space that includes most important MOs participating in physical or chemical reactions is an effective approach to form compact representations of molecular orbital spaces of a given basis set. Properly constructed active space reduces the number of MOs used, and is critical for performing quantum computations for large systems in the NISQ era; however, 
\response{a comprehensive method that can select an appropriate active space for correlated molecular calculations, regardless of the basis sets, is still desirable. For example, a conventional active space selection scheme based on the occupation numbers of nature orbitals often identifies important orbitals successfully, but they are not foolproof and might falter in specific scenarios \cite{Khedkar:JCTC2019}. Therefore it remains worth inspecting suitable ways to choose an active space for the system under consideration.}
%a standard protocol that can consistently select a favorable active space for correlated molecular calculations in a given basis set remains unavailable. Therefore, a proper way to choose an active space is worth inspecting.

%\textcolor{orange}{\sout{While a real-space basis set provides highly accurate calculations, we further inspect on the correlated wave function method in the Daubechies wavelet basis set and find that the approach of using the MB set of Daubechies wavelet MOs \cite{Hong:PRXQ2022} does not provide adequate results of vibrational frequencies as it fails to produce smooth potential energy curves (PECs) for some molecules, and furthermore by excluding those unavailable data, it also has a significantly larger root-mean-square deviation (RMSD) value even though the number of active MOs used is considerably higher than that of our proposed approach (see Supplemental Material \cite{SupplementalMaterial}).}} }

In addition, 
%the electron correlation effect could be incorporated into the basis set via the exchange-correlation (XC) functional by using the MOs derived from the Kohn-Sham (KS) density functional theory (DFT) \cite{Grimme:CPL1996,Grimme:JCP1999,Bour:CPL2001,Veseth:JCP2001,Beran:PCCP2003,Kim:PCCP2015,Kim:JCP2016,Bertels:JCTC2021}.
\response{several studies have demonstrated that using the MOs derived from the Kohn-Sham (KS) density functional theory (DFT) in correlated \pdfproof{wave function} theory calculations could provide improved results. \cite{Grimme:CPL1996,Grimme:JCP1999,Bour:CPL2001,Veseth:JCP2001,Beran:PCCP2003,Kim:PCCP2015,Kim:JCP2016,Bertels:JCTC2021} 
For example, Bertels \pdfproof{\textit{et al.}~}\cite{Bertels:JCTC2021} have recently benchmarked classical CCSD(T) calculations with a variety of KS MOs on many diatomic molecules. In their paper, MOs derived from a hierarchy of different exchange-correlation (XC) functionals were compared and an improvement on predicting vibrational frequencies was observed. Their results indicate that the electron correlation effect could be incorporated into the MOs via the XC functional, which} suggests that a possible further improvement on the accuracy of molecular vibrational frequency calculations could be realized when using KS MOs without increasing the number of MOs used.
In this work, we propose a quantum computational approach that adopts the molecular orbitals derived from
KS DFT with XC functionals and expanded in the Daubechies wavelet basis set, where a reduced active space based on an energy criterion of 
%\sout{a second-order Møller–Plesset perturbation theory (MP2) \cite{MP2}}
a first-order pair energy in the theory of independent electron pair approximation (IEPA) \cite{Sinanogu:ACP1964,Nesbet:ACP1965}, denoted by IEPA1, is selected to further reduce the required number of qubits. 
We perform a VQE quantum computing benchmark investigation using our proposed approach on the harmonic vibrational frequencies of 43 neutral, closed-shell diatomic molecules with results in great agreement with their corresponding experimental data.
We attribute its excellent performance to three factors:
(i) a better description of the Hamiltonian by the Daubechies wavelets MOs, 
(ii) better reference for the electron correlation effect in the MOs via the XC functional, 
(iii) an improved selection of active space by IEPA1 energy criteria.
Remarkably, our proposed approach significantly reduces the number of qubits 
for the 43 diatomic molecules from 20 to 60 using the traditional cc-pVDZ basis set with frozen core approximation to only 2 to 12 but with similar accuracy for the obtained results.  
VQE calculations by our approach with a  significantly reduced qubit number imply that a considerably reduced 
ansatz circuit depth can reach the same level of accuracy as those by using the traditional cc-pVDZ basis set.
For example, the VQE calculation of a H$_2$ molecule using the cc-pVDZ basis set requires 20 qubits and the circuit depth estimated by using the UCCSD ansatz would be, as mentioned above, over $10^4$, and even for the case of using the heuristic ansatz of a RealAmplitudes  circuit,  to have a result reaching the same level of accuracy
the estimated circuit depth would be still about a few hundreds.  
In contrast, the required circuit depth of our proposed approach to obtain a result of harmonic vibrational frequency of H$_2$ in great agreement with the experimental data is only 4. 
For all the 43 molecules we investigate, the required circuit depths to reach accurate vibrational frequency results are all less than 100. Thus, VQE quantum computations of these molecules using our proposed approach are promisingly realizable on the NISQ devices given the recent advance on quantum utility demonstration \cite{Kim:Nature2023}  and
the projected achievement of the so-called $100\times100$ Challenge in 2024 
\cite{IBMQ:100x100}.

%We find that the approach of using the MB set of Daubechies wavelet MOs \cite{Hong:PRXQ2022} does not provide adequate results of vibrational frequencies as it fails to produce smooth potential energy curves (PECs) for some of the 43 molecules, and furthermore by excluding those unavailable data, it also has a significantly larger root-mean-square deviation (RMSD) value even though the number of active MOs used is considerably higher than that of our proposed approach (see Supplemental Material \cite{SupplementalMaterial}).
%adopts Daubechies Wavelet basis set and a selected active space using the for the calculation of harmonic vibrational frequency. 

The paper is organized as follows. In Sec.~\ref{sec:Method}, we introduce the methods we use to calculate the molecular vibrational frequencies in our proposed approach. The active space selection criterion by IEPA1 will be described in
Sec.~\ref{subsec:Active_Space}. In Sec.~\ref{subsec:EDQC},
the performance of the proposed approach is benchmarked against the classical WFT methods and traditional basis sets on a large dataset of 43 neutral closed-shell diatomic molecules. Moreover, we use VQE to evaluate the performance of the UCCSD ansatz with the Hamiltonian represented by the selected Daubechies wavelet basis set in Sec.~\ref{subsec:UCCSD}. The results show that the UCCSD ansatz can yield accurate results except for systems whose Mayer bond order indices \cite{Mayer:CPL1983} are larger than 2. For those systems, we then demonstrate that the heuristic hardware-efficient ansatz even with a substantially shorter circuit depth can provide significant improvement over the UCCSD ansatz in Sec.~\ref{subsec:AnsatzCf}. To the best of our knowledge, our investigation is the first systematical benchmark study to demonstrate that a heuristic hardware-efficient ansatz could outperform a chemistry-inspired UCCSD ansatz in predicting accurate molecular properties by quantum computation. Such a comprehensive benchmark study enables us to establish an accurate approach for the vibrational frequencies of diatomic molecules, which could be realized on near-term NISQ computers. A conclusion and the outlook of this work are presented in Sec.~\ref{sec:Conclusion}.

%------------------------------------------------------------------------------
%  METHODS
%------------------------------------------------------------------------------
\section{METHODS} \label{sec:Method}
\subsection{Daubechies \pdfproof{w}avelet}
In this work, MOs expanded in the Daubechies wavelet basis set \cite{Daubechies} are generated from the \pdfproof{\textsc{bigdft}} package \cite{Genovese:JCP2008,Genovese:CRM2011,Mohr:JCP2014,Ratcliff:JCP2020}.
The KS MO is expanded in the Daubechies wavelets of order 16 with one scaling function $\phi^0$ and seven augmented wavelets $\psi^1,\cdots,\psi^7$ \cite{Genovese:JCP2008,Hong:PRXQ2022}:
%\small
%\begin{equation}
%  \Psi^{\text{KS}}(\mathbf{r}) = \sum_{i_1,i_2,i_3} s_{i_1,i_2,i_3} \ \phi_{i_1,i_2,i_3}(\mathbf{r}) + \sum_{j_1,j_2,j_3} \sum_{\nu=1}^7 w_{j_1,j_2,j_3}^\nu \ \psi^{\nu}_{j_1,j_2,j_3}(\mathbf{r}),
%\end{equation}
%\normalsize
\begin{equation}
  \Psi^{\text{KS}}(\mathbf{r}) = \sum_{\mathbf{i}} s_{\mathbf{i}} \ \phi_{\mathbf{i}}^0(\mathbf{r}) + \sum_{\mathbf{j}} \sum_{\lambda=1}^7 w_{\mathbf{j}}^\lambda \ \psi_\mathbf{j}^\lambda(\mathbf{r}),
\end{equation}
where $s_{\mathbf{i}}, w_{\mathbf{j}}$ are expansion coefficients, and indices $\mathbf{i}=\{i_1,i_2,i_3\}$ and $\mathbf{j}=\{j_1,j_2,j_3\}$ are summed over the low (coarse) and high (fine) resolution regions, respectively, in three-dimensional real space grid points $\mathbf{r}=\{x,y,z\}$ with grid spacing $h$.
The three-dimensional basis functions $\phi^0_{i_1,i_2,i_3}(\mathbf{r})$ and $\phi^\lambda_{j_1,j_2,j_3}(\mathbf{r})$ are a tensor product of one-dimensional scaling function $\phi$ and wavelet $\psi$, which read as
\begin{align}
  \phi^0_{i_1,i_2,i_3}(\mathbf{r}) &= \phi(x/h-i_1) \phi(y/h-i_2) \phi(z/h-i_3), \nonumber\\
  \psi^1_{j_1,j_2,j_3}(\mathbf{r}) &= \psi(x/h-j_1) \phi(y/h-j_2) \phi(z/h-j_3), \nonumber\\
  \psi^2_{j_1,j_2,j_3}(\mathbf{r}) &= \phi(x/h-j_1) \psi(y/h-j_2) \phi(z/h-j_3), \nonumber\\
  \psi^3_{j_1,j_2,j_3}(\mathbf{r}) &= \psi(x/h-j_1) \psi(y/h-j_2) \phi(z/h-j_3), \nonumber\\
  \psi^4_{j_1,j_2,j_3}(\mathbf{r}) &= \phi(x/h-j_1) \phi(y/h-j_2) \psi(z/h-j_3), \nonumber\\
  \psi^5_{j_1,j_2,j_3}(\mathbf{r}) &= \psi(x/h-j_1) \phi(y/h-j_2) \psi(z/h-j_3), \nonumber\\
  \psi^6_{j_1,j_2,j_3}(\mathbf{r}) &= \phi(x/h-j_1) \psi(y/h-j_2) \psi(z/h-j_3), \nonumber\\
  \psi^7_{j_1,j_2,j_3}(\mathbf{r}) &= \psi(x/h-j_1) \psi(y/h-j_2) \psi(z/h-j_3).
\end{align}
The multiresolution of the Daubechies wavelets of order 16 is featured by the refinement equations
\begin{align}
  \phi(x) = \sqrt{2} \sum_{l=-7}^8 h_l \phi(2x-l), \nonumber\\
  \psi(x) = \sqrt{2} \sum_{l=-7}^8 g_l \phi(2x-l),
\end{align}
which establishes a relation between the scaling functions on a twice coarser grid and a finer grid. The coefficients $h_l$ and $g_l=(-1)^l h_{-l}$ are filters that characterizes the scaling function and wavelet.

In a simulation domain, the chemical bonds are described in a high resolution region (fine region) which is composed of one scaling function and seven wavelets, and the exponentially decaying tails of the \pdfproof{wave functions} are described in a low resolution region (coarse region) which is only composed of scaling functions.

The number of virtual orbitals is a parameter and it is chosen to be equal to the total number of atomic input orbitals of the system as implemented in the \pdfproof{\textsc{bigdft}} package, while degenerate orbitals will be considered together. This forms the initial truncated MO space. The spin treatment does not involve spin polarization. 
The XC functionals considered in this work are Hartree-Fock (HF), Perdew-Burke-Ernzerhof (PBE) \cite{PBE}, and PBE0 \cite{PBE0}. 
Other popular XC functionals which have no suitable pseudopotential in the package, like B3LYP \cite{B3LYP}, are excluded. All the three XC functionals use the same norm-conserving Hartwigsen-Goedeker-Hutter Krack (HGH-K) \cite{GTH:PRB1996, HGH:PRB1998, Krack:TCA2005} pseudopotential generated with the PBE functional. 

One caveat needed to be mentioned in this work is about the grid parameters in \pdfproof{\textsc{bigdft}}. The grid parameters, hgrids which controls the grid spacing and rmult which controls the size of simulation space, are determined from the analysis of the grid convergence for each molecule.
However, such determination is analyzed in the framework of DFT where virtual orbitals are not used, and thus might not be optimal for WFT since these grid parameters significantly affect the properties such as the shapes and energies of the virtual orbitals. Generally, setting better grid parameters (smaller hgrids and larger rmult) helps the calculation converge to a lower energy but with increasing computing cost, so the optimal choice should consider both accuracy and efficiency. The setting with better and better grid parameters could generate \pdfproof{continuumlike} orbitals, but WFT with \pdfproof{continuumlike} virtual orbitals would suffer from the vanishing electron correlation \cite{Natarajan:CP2012}. Therefore to find a better way to determine the grid parameters for WFT is crucial and will be investigated further in the future.

\subsection{Second-\pdfproof{q}uantized Hamiltonian}
Given a set of KS MOs, the second-quantized Hamiltonian is constructed as
\begin{equation}
  H = \sum_{p,q} h_{pq} a_p^\dag a_q + \frac{1}{2} \sum_{p,q,r,s} h_{pqrs} a_p^\dag a_q^\dag a_r a_s,
\end{equation}
where the one-electron integral
\small
\begin{equation}
  h_{pq} = \int d\mathbf{x} \ \Psi^{\text{KS}^*}_p(\mathbf{x}) \left( -\frac{\nabla^2}{2} - \sum_A \frac{Z_A}{|\mathbf{r} - \mathbf{R}_A |} \right) \Psi^{\text{KS}}_q(\mathbf{x}),
\end{equation}
\normalsize
the two-electron integral
\small
\begin{equation}
  h_{pqrs} = \int d\mathbf{x}_1 d\mathbf{x}_2 \frac{\Psi^{\text{KS}^*}_p(\mathbf{x}_1) \Psi^{\text{KS}^*}_q(\mathbf{x}_2)  \Psi^{\text{KS}}_r(\mathbf{x}_2) \Psi^{\text{KS}}_s(\mathbf{x}_1) }{|\mathbf{r}_1-\mathbf{r}_2|},
\label{2eintegral}  
\end{equation}
\normalsize
and $a_p^\dag$ and $a_q$ are creation and annihilation operators acting on the $p$th and $q$th component of the occupation number vector in Fock space, respectively. The $\mathbf{x} = (\mathbf{r},\sigma)$ denotes the position and the spin of the electron, and $\mathbf{R}_A$ and $Z_A$ denote the position and the atomic number of the $A$th nucleus, respectively. The values of the integrals $h_{pq}$ and $h_{pars}$ are calculated via \pdfproof{\textsc{bigdft}} subroutines.

% Correlation Energy
The correlation energy, $E_{\text{corr}}$, defined in WFT with KS MOs is \cite{Beran:PCCP2003}
\begin{equation}
  E_{\text{corr}} = E_{\text{exact}} - \langle \Psi^{\text{KS}} | H | \Psi^{\text{KS}} \rangle,
\end{equation}
where $E_{\text{exact}}$ is the exact energy, $|\Psi^{\text{KS}} \rangle$ is the single Slater determinant formed with a set of KS spin orbitals and $\langle \Psi^{\text{KS}} | H | \Psi^{\text{KS}} \rangle$ is the reference value mimicking the HF energy.

% Active Space
\subsection{Active \pdfproof{s}pace} \label{subsec:Active_Space}
We discuss here how the [MB] active space and [IEPA1] active space are selected.
The [MB] active space selected from the initial truncated MO space is chosen to imitate the complete active space constructed from the minimal basis set. Let us take LiF as an example. The minimal basis set for this system is constructed from the \pdfproof{$1s$}, \pdfproof{$2s$}, and \pdfproof{$2p$} orbitals of Li, and the \pdfproof{$1s$}, \pdfproof{$2s$}, and \pdfproof{$2p$} orbitals of F, giving a total of 10 MOs from a linear combination of atomic orbitals (LCAO).
Since this system has 12 electrons resulting in 6 occupied MOs, one would take the 6 occupied MOs and 4 lowest-energy unoccupied MOs in the [MB] method to select a total of 10 MOs in the active space.
However, the usual frozen core approximation is used here for F’s \pdfproof{$1s$} orbital, but we do not apply this approximation to Li’s \pdfproof{$1s$} orbital since the pseudopotential used in this work includes that orbital (the same treatment applied to the alkaline earth metal element Be, while for Na atom the 1s orbital is frozen). 
By taking these considerations into account, the active space of LiF 
consists of 10 active electrons and 9 active MOs, denoted as [10,9], in our [MB] approach. Note that for \pdfproof{g}roup 13 - \pdfproof{g}roup 17 atoms in the periodic table of the elements, the inner-shell frozen core approximation is used and the total number of minimal-basis valence orbitals for these atoms is 4.
%$2^{\text{nd}}$-, $3^{\text{rd}}$-, and $4^{\text{th}}$-period 

The active orbitals in the [IEPA1] active space are selected from the initial truncated MO space by calculating the first-order pair energy in the theory of the independent electron pair approximation. In the following, we describe briefly the theory of IEPA and then present an energy criterion derived from IEPA1 to select active orbitals. The theory of IEPA considers the correlation energy associated with a pair of electrons independently of other pairs in a configuration interaction way. The correlated \pdfproof{wave function} for the pair $ij$, denoted as the pair function $|\Psi_{ij}\rangle$, is
\begin{align}
    |\Psi_{ij}\rangle = |\Psi_0\rangle + \sum_{a<b} c_{ij}^{ab} |\Psi_{ij}^{ab}\rangle,
\end{align}
where \pdfproof{$i$ and $j$} denote the occupied spin orbital indices, \pdfproof{$a$ and $b$} denote the virtual spin orbital indices\pdfproof{,} and $c_{ij}^{ab}$ is the \pdfproof{wave function} coefficient. Here,
$|\Psi_0\rangle$ and $|\Psi_{ij}^{ab}\rangle$ are ground and doubly excited Slater determinants formed with a set of HF spin orbitals.
The energy of this correlated \pdfproof{wave function}, denoted by $E_{ij}^{\text{IEPA}}$, is
\begin{align}
    E_{ij}^{\text{IEPA}} = \langle \Psi_{ij} | H | \Psi_{ij} \rangle = E_0 + e_{ij}^{\text{IEPA}}\pdfproof{,}
\end{align}
where $E_0$ is the HF reference energy and $e_{ij}^{\text{IEPA}}$ is the pair (correlation) energy. Under the first-order approximation to IEPA, which neglects coupling between excited determinants, the first-order pair energy \cite{Szabo}, denoted by $e_{ij}^{\text{IEPA1}}$, reads 
\begin{align}
    e_{ij}^{\text{IEPA1}} = \sum_{a<b}^{\text{vir}} \frac{|\langle \Psi_0 |H| \Psi_{ij}^{ab} \rangle|^2}{\epsilon_i + \epsilon_j-\epsilon_a-\epsilon_b} 
    = \sum_{a<b}^{\text{vir}} \frac{\left(h_{ijab}-h_{ijba}\right)^2}{\epsilon_i + \epsilon_j-\epsilon_a-\epsilon_b},
\end{align}
%and approximates the energy difference between $|\Psi_0\rangle$ and $|\Psi_{ij}^{ab}\rangle$ by the difference of orbital energies
where $h_{ijab}$ is the  two-electron integral defined in Eq.~(\ref{2eintegral}) and the energy difference in the denominator, $\langle\Psi_{ij}^{ab} | H - E_0 |\Psi_{ij}^{ab}\rangle$, 
%between $|\Psi_0\rangle$ and $|\Psi_{ij}^{ab}\rangle$ in the denominator 
has been approximated by the difference of orbital energies, $\epsilon_i + \epsilon_j-\epsilon_a-\epsilon_b$.
The total first-order pair energy is then
\begin{align}
    E_\text{IEPA1} &= \sum_{i<j}^{\text{occ}} e_{ij}^{\text{IEPA1}} 
    = \sum_{i<j}^{\text{occ}} \sum_{a<b}^{\text{vir}} \frac{\left(h_{ijab}-h_{ijba}\right)^2}{\epsilon_i + \epsilon_j-\epsilon_a-\epsilon_b} \\
    &= E_\text{MP2}\pdfproof{,}
\end{align}
which is identical to the energy correction by the second-order M\o ller–Plesset perturbation theory (MP2), denoted by $E_\text{MP2}$ \cite{MP2}.

The pair energy would suggest a selection criterion for the active space.  The active orbitals of the [IEPA1] active space are selected from the MOs in initial truncated MO space by using the first-order pair energy.
In order to determine which MO is important, we calculate $E_{\text{IEPA1}}$ of the MO by considering the sum of all pair energies involved that MO, that is, the terms in the summation involving only the spin indices of that MO.
For example, the first occupied MO is denoted as OccMO[0], and its spin-up and spin-down orbitals are labeled with indices $i=0$ and $i=1$, respectively; in this case, $E_{\text{IEPA1}}$ of OccMO[0] is
\begin{align}
  &E_{\text{IEPA1}} (\text{OccMO[0]}) = \left( \sum_{i=0<j}^{\text{occ}} + \sum_{i=1<j}^{\text{occ}} \right) e_{ij}^{\text{IEPA1}} \nonumber \\
  &=\left( \sum_{i=0<j}^{\text{occ}} + \sum_{i=1<j}^{\text{occ}} \right) \sum_{a<b}^{\text{vir}} \frac{\left(h_{ijab}-h_{ijba}\right)^2}{\epsilon_i + \epsilon_j-\epsilon_a-\epsilon_b}. 
\end{align}
Take each $E_{\text{IEPA1}}$ of the MO divided by the total $E_{\text{IEPA1}}$ as a percentage, and then with a target of selecting a small number of MOs, the MOs with relatively large percentages are chosen into the IEPA1 active space.
A different flavor to directly determine MRA-represented pair-natural orbitals on the level of MP2 can be found in \pdfproof{Refs.~}\cite{Kottmann:JCP2020, Kottmann:JPCL2021}; the number of active pair-natural orbitals (approximate nature orbitals), is truncated based on occupation numbers. Compared to their approach \cite{Kottmann:JCP2020, Kottmann:JPCL2021}, our approach directly analyzes on canonical orbitals without additional transformation to nature orbitals.

While there is no distinguishable difference between the energies $E_{\text{IEPA1}}$ and $E_{\text{MP2}}$ for the standard HF orbitals, we follow by the idea of IEPA to evaluate the first-order pair energy when the KS orbitals are used. However, in this case the Brillouin's theorem, which states that the matrix element contributed from singly excited Slater determinants formed with a set of HF spin orbitals is zero, does not hold due to the fact that the KS MO is not the eigenfunction of the Fock operator.

% Quantum Computing
\subsection{Quantum \pdfproof{c}omputing}
Quantum computing in VQE starts from mapping the second-quantized Hamiltonian to the qubit Hamiltonian, and in this work the quantum computing package, \pdfproof{\textsc{qiskit}} \cite{Qiskit}, is used. 
Common encoding methods to encode the fermionic operators to qubit operators transform a fermionic system of $m$ active MOs ($2m$ spin orbitals) into an $2m$-qubit system.
Here, the parity encoding scheme \cite{Bravyi:2002AP,Seeley:JCP2012} is chosen to further reduce the number of qubits by two due to $\mathbb{Z}_2$ symmetry reduction \cite{Bravyi:arXiv2017}, so the number of qubits used is $2m-2$.

The ground state energy of the qubit Hamiltonian is calculated via different methods.
In Sec.~\ref{subsec:EDQC}, the exact diagonalization of the qubit Hamiltonian is used and we denote this approach as the exact diagonalization method of quantum computing (EDQC). Hence, the results of the EDQC method could be regarded as the best results achievable by quantum computation. 
Consequently, the EDQC method is also used as the standard to investigate the performance of the VQE method in \pdfproof{Secs.~}\ref{subsec:UCCSD} and \ref{subsec:AnsatzCf}, where we demonstrate the result of the vibrational frequencies obtained by our proposed approach for VQE quantum computation is as accurate as those by the EDQC method.
At the same time, the state fidelity between the EDQC ground state and the VQE ground state calculated via different circuit ansatzes is evaluated as another verification indicator (see Table \ref{tab:AnsatzComparison}).

The quantum circuit ansatzes used here for VQE are the UCCSD ansatz and the heuristic hardware-efficient ansatz of RealAmplitudes from the \pdfproof{\textsc{qiskit}} circuit library, where these VQE approaches are denoted as VQE(UCCSD) and VQE(RealAmplitudes), respectively. The Hartree-Fock state as the initial state is prepended to both the quantum circuits. The entanglement type for the RealAmplitudes ansatz is chosen to be the linear entanglement. The optimizers used are SLSQP and L-BFGS-B of \pdfproof{\textsc{scipy}} \cite{SciPy}. All the VQE calculations are performed in the noiseless situation with the state-vector simulation method.

% Classical Computing
\subsection{Classical \pdfproof{c}omputing}
We use the \pdfproof{\textsc{pyscf}} package \cite{PySCF} for the classical computations. The methods used include CCSD(T) and complete active space configuration interaction (CASCI) \cite{CASCI}, where the former is chosen since it is the golden standard method in quantum chemistry and the latter is chosen since it is the FCI on the active space which corresponds to the exact diagonalization method in this work. The KS MOs with XC functional of HF or PBE0 are used, and the nature orbital \cite{NO} is additionally considered in the CASCI method.
The traditional basis set used is the Dunning correlation-consistent basis set, cc-pVDZ. The spin treatment is spin-restricted Hartree-Fock method. The number of frozen cores is chosen to be the same as that of the pseudopotential considered in this work. In CASCI, for a rapid convergence, the MP2 natural orbitals, denoted by MP2NO, transformed from the standard HF MOs are used and then the active space is determined by the natural orbital occupation number, denoted by NOON. 

In the calculation of Mayer bond order indices, we use \pdfproof{\textsc{pyscf}} functions to evaluate the formula. 
This index between atoms A and B of a closed-shell molecule is defined as
\begin{equation}
  M_{AB} = \sum_{\mu \in A} \sum_{\nu \in B} (\mathbf{DS})_{\mu\nu} (\mathbf{DS})_{\nu\mu},   
\end{equation}
where $\mu,\nu$ are indices for the basis functions belonged to the assigned atom, and $\mathbf{DS}$ denotes the product the spinless density matrix $\mathbf{D}$ and the overlap matrix $\mathbf{S}$.

% Harmonic Vibrational Frequency
\subsection{Harmonic \pdfproof{v}ibrational \pdfproof{f}requency}
The quantity to be calculated in the benchmark is the harmonic vibrational frequency. \response{For diatomic molecules, there is only one vibrational mode, the stretching mode. The corresponding diatomic harmonic vibrational frequency} is calculated by quadratic polynomial curve fitting using five points around the minimal energy point of the equilibrium bond length with step size 0.01 \text{\r{A}} on the PEC. %potential energy curve (PEC).  
In order to put different comparative methods on equal footing, this calculation procedure applies to all the methods considered here. The equilibrium bond lengths calculated by different methods are presented in \pdfproof{Ref.~}\cite{SupplementalMaterial}.

% DataSet
\subsection{Data\pdfproof{s}et} \label{sec:dataset}
For simplicity, the dataset considered consists of diatomic molecules that are neutral, closed-shell and formed by atoms (elements) in row 1 to row 4 of the periodic table, excluding the transition metal elements, but the diatomic molecules whose experimental data are not available on the Computational Chemistry Comparison and Benchmark Database \cite{CCCBDB} are also not considered. For comparison, since there is no K atom in the cc-pVDZ basis set, the molecules involved K atom are excluded.
In \pdfproof{\textsc{bigdft}}, we can not generate smooth PECs for NaLi, Na$_2$ and NaK, and therefore these three molecules are excluded. C$_2$ with multireference character and F$_2$ owing to severe static correlation \cite{Martin:JPCA2015} are taken to be the overall outliers and then are excluded. In the end, the benchmark dataset contains 43 neutral closed-shell diatomic molecules.

% Notation
\subsection{Notations}
We use the following notations to denote different 
approaches used in this work: ``Method[active space selection method]-XC/Basis Set'', where XC can be HF, PBE, or PBE0 to keep track of the type of MOs. For example,
EDQC[IEPA1]-PBE0/Wavelet, denotes using the EDQC method with PBE0 exchange functional for the Daubechies Wavelet MO basis set, and the method of the active space selection is IEPA1, where Wavelet is shorthand for Daubechies Wavelet MO basis set.  

The notation [MB] or [IEPA1] indicates how the active space are selected as we discuss in Sec.~\ref{subsec:Active_Space}. The active space indices $[n,m]$ in Table \ref{tab:Benchmark_HVF} denote $n$ active electrons and $m$ active MOs ($2m$ spin orbitals). 
%For methods like CCSD(T)-XC/cc-pVDZ without introducing the active space, indices $[n,m]$ are used to represent the number of the electrons, $n$, and the number of MOs, $m$, of the full size of the basis set, which can be compared to the number of MOs used in different approaches (see Table \ref{tab:Benchmark_HVF}).

%------------------------------------------------------------------------------
%  RESULTS
%------------------------------------------------------------------------------

\begin{table*}[htp]
  \centering
  \rotatebox{90}{
  \begin{varwidth}{\textheight}  
  \caption{Harmonic vibrational frequencies (in cm$^{-1}$) of neutral closed-shell diatomic molecules obtained by different methods.\hfill} %\fhill to make caption left aligned
  \renewcommand{\arraystretch}{0.92}% Tighter
  \begin{ruledtabular}
    \begin{tabular}{lrrrrrrrrr}
         &         EDQC  &         EDQC  &         EDQC  &         CASCI  &         CASCI  &        CASCI  &  CCSD(T)          & CCSD(T)       &        \\
         &   [IEPA1]-HF  &  [IEPA1]-PBE  & [IEPA1]-PBE0  &  [IEPA1]-PBE0  &     [MB]-PBE0  & [NOON]-MP2NO  &   -PBE0           &    -HF        &        \\
Mol.     &     /Wavelet  &     /Wavelet  &     /Wavelet  &      /cc-pVDZ  &      /cc-pVDZ  &     /cc-pVDZ  &/cc-pVDZ           &/cc-pVDZ       &   Expt.\\
\hline%----------------------------------------------------------------------------------------------------------------------------------------\\
H$_2$    & [2,2]4462.40  & [2,2]4365.33  & [2,2]4391.45  &  [2,3]4194.85  &  [2,2]4291.21  & [2,2]4224.68  &[2,10]4397.89     &[2,10]4397.70   &4401.21\\
LiH      & [2,3]1413.65  & [2,3]1393.42  & [2,3]1405.36  &  [2,5]1309.67  &  [4,6]1372.62  & [2,5]1393.25  &[4,19]1350.18     &[4,19]1350.18   &1405.50\\
NaH      & [2,3]1178.10  & [2,3]1149.01  & [2,3]1143.76  &  [2,5]1153.83  & [10,9]1107.69  & [2,5]1128.44  &[12,23]1107.59    &[12,23]1107.47   &1171.97\\
BH       & [4,5]2476.84  & [4,7]2418.72  & [4,7]2386.09  &  [4,9]2353.90  &  [4,5]2166.48  & [4,9]2390.04  &[6,19]2318.76     &[6,19]2319.91   &2366.73\\
AlH      & [4,6]1707.87  & [4,7]1756.43  & [4,7]1709.33  &  [4,9]1649.35  &  [4,5]1554.54  & [4,9]1666.97  &[14,23]1672.14    &[14,23]1672.42   &1682.38\\
GaH      & [4,6]1600.18  & [4,7]1575.06  & [4,7]1593.80  &  [4,8]1600.89  &  [4,5]1509.07  & [4,9]1612.96  &[32,32]1602.76    &[32,32]1597.07   &1604.52\\
HF       & [6,6]4432.96  & [2,3]4152.30  & [2,3]4148.44  & [8,10]4631.00  &  [8,5]4141.97  & [6,6]4171.69  &[10,19]4151.90    &[10,19]4151.79   &4138.39\\
HCl      & [2,3]3008.83  & [2,3]2883.96  & [2,3]2874.66  &  [8,9]2900.94  &  [8,5]2858.48  & [8,9]3162.91  &[18,23]3018.33    &[18,23]3017.70   &2990.93\\
HBr      & [2,4]2623.92  & [2,3]2515.62  & [2,3]2511.84  &  [8,9]2615.01  &  [8,5]2541.63  & [8,9]2710.44  &[36,32]2654.50    &[36,32]2654.02   &2649.00\\
LiF      & [4,6]\, 932.22  & [4,6]\, 939.94  & [4,6]\, 938.00  & [8,7]\, 959.51& [10,9]\, 999.36 &[8,7]\, 980.57  &[12,28]\, 975.01  &[12,28]\, 975.36   & 910.57\\
LiCl     & [4,6]\, 648.66  & [4,4]\, 663.29  & [4,4]\, 649.26  & [8,7]\, 656.13& [10,9]\, 632.17 &[6,6]\, 635.55  &[20,32]\, 625.11  &[20,32]\, 624.96   & 642.95\\
LiBr     & [6,5]\, 558.04  & [6,5]\, 571.19  & [6,5]\, 559.84  & [8,7]\, 540.32& [10,9]\, 555.71 &[6,8]\, 582.25  &[38,41]\, 554.50  &[38,41]\, 554.87   & 563.00\\
NaF      & [6,5]\, 541.11  & [6,4]\, 552.30  & [6,4]\, 548.70  & [8,7]\, 589.84&[16,12]\, 594.09 &[6,6]\, 584.83  &[20,32]\, 582.15  &[20,32]\, 582.36   & 535.66\\
NaCl     & [6,4]\, 362.28  & [6,4]\, 363.57  & [6,4]\, 362.88  & [8,7]\, 373.37&[16,12]\, 354.34 &[6,6]\, 343.89  &[28,36]\, 349.63  &[28,36]\, 349.83   & 364.68\\
NaBr     & [6,4]\, 292.53  & [6,4]\, 293.19  & [6,4]\, 292.99  & [8,7]\, 354.87&[16,12]\, 289.84 &[6,8]\, 311.71  &[46,45]\, 288.17  &[46,45]\, 288.17   & 302.00\\
BeO      & [6,9]1506.27  & [6,6]1322.15  & [6,6]1436.71  &  [8,6]1635.71  & [10,9]1262.73  & [6,6]1500.43  &[12,28]1365.29    &[12,28]1358.46   &1457.09\\
BeS      & [6,8]\, 866.53  & [6,6]1002.45  & [6,6]\, 949.56   &  [8,6]1085.78 &[10,9]\, 911.52 &[6,6]\, 984.06  &[20,32]\, 960.90  &[20,32]\, 959.81   & 997.94\\
BF       & [6,7]1353.21  & [8,7]1495.16  & [8,7]1462.33  & [10,8]1504.25  & [10,8]1277.04  & [8,7]1349.93  &[14,28]1304.94    &[14,28]1306.53   &1402.16\\
BCl      & [8,6]\, 859.42  & [8,6]\, 873.82  & [8,6]\, 852.18  &[10,8]\, 772.05 &[10,8]\, 772.15 &[8,7]\, 867.14  &[22,32]\, 841.81  &[22,32]\, 840.77   & 840.29\\
BBr      & [8,6]\, 676.96  & [8,6]\, 686.53  & [8,6]\, 686.03  &[10,8]\, 644.16 &[10,8]\, 644.07 &[8,7]\, 700.09  &[40,41]\, 695.24  &[40,41]\, 694.65   & 684.31\\
AlF      & [8,7]\, 815.15  & [8,7]\, 839.01  & [8,7]\, 814.55  &[10,8]\, 791.47 &[10,8]\, 787.41 &[8,7]\, 777.61  &[22,32]\, 788.84  &[22,32]\, 789.40   & 802.32\\
AlCl     & [2,3]\, 480.17  & [4,4]\, 486.10  & [4,4]\, 474.55  &[10,7]\, 455.72 &[10,8]\, 438.92 &[8,7]\, 462.90  &[30,36]\, 474.84  &[30,36]\, 474.51   & 481.77\\
AlBr     & [2,3]\, 374.09  & [4,4]\, 375.63  & [4,4]\, 371.20  &[10,7]\, 363.31 &[10,8]\, 344.91 &[8,7]\, 374.39  &[48,45]\, 375.61  &[48,45]\, 375.12   & 378.00\\
GaF      & [8,6]\, 591.41  & [8,6]\, 623.13  & [8,6]\, 609.27  &[10,8]\, 641.78 &[10,8]\, 646.70 &[8,7]\, 654.89  &[40,41]\, 659.63  &[40,41]\, 659.25   & 622.20\\
GaCl     & [4,4]\, 333.49  & [4,4]\, 353.81  & [4,4]\, 340.81  &[10,7]\, 356.66 &[10,8]\, 340.84 &[8,7]\, 365.68  &[48,45]\, 389.50  &[48,45]\, 372.43   & 365.30\\
CO       & [8,8]2126.40  & [8,6]2297.41  & [8,6]2248.86  & [10,8]2415.09  & [10,8]2118.43  & [8,7]2189.97  &[14,28]2101.17    &[14,28]2180.17   &2169.76\\
CS       & [8,6]1334.09  & [8,7]1338.66  & [8,7]1298.67  & [10,8]1185.46  & [10,8]1185.46  & [8,7]1282.47  &[22,32]1248.77    &[22,32]1250.42   &1285.16\\
CSe      &[10,7]1086.94  & [8,8]\, 903.59  & [8,7]1010.42  &[10,8]\, 964.48 &[10,8]\, 963.57  & [8,7]1036.65  &[40,41]1006.64    &[40,41]1007.86   &1035.36\\
SiO      & [8,8]1311.41  & [8,7]1238.18  & [8,7]1247.22  & [10,7]1239.35  & [10,8]1100.60  & [8,7]1200.23  &[22,32]1139.71    &[22,32]1140.03   &1241.54\\
SiS      & [8,8]\, 784.94  & [8,8]\, 732.18  & [8,8]\, 716.80  &[10,8]\, 682.38 &[10,8]\, 682.32 &[8,7]\, 732.23  &[30,36]\, 727.85  &[30,36]\, 727.80   & 749.65\\
SiSe     & [8,6]\, 554.84  & [8,6]\, 591.19  & [8,6]\, 582.23  &[10,8]\, 538.16 &[10,8]\, 538.59 &[8,7]\, 571.14  &[48,45]\, 569.19  &[48,45]\, 569.55   & 580.00\\
GeO      & [8,8]1074.76  & [8,6]\, 995.56  & [8,6]\, 999.61  &[10,8]\, 885.21 &[10,8]\, 885.82 &[8,7]\, 989.61  &[40,41]\, 939.51  &[40,41]\, 957.82   & 985.50\\
N$_2$    & [4,4]2345.03  & [4,4]2400.36  & [4,4]2399.42  & [10,8]2636.51  & [10,8]2146.61  & [8,7]2332.91  &[14,28]2329.44    &[14,28]2331.16   &2358.57\\
PN       & [4,4]1365.63  & [4,4]1375.26  & [4,4]1375.98  & [10,8]1220.52  & [10,8]1220.52  & [8,7]1301.49  &[22,32]1254.95    &[22,32]1293.57   &1336.95\\
P$_2$    & [4,4]\, 800.32  & [4,4]\, 796.93  & [4,4]\, 778.53  &[10,8]\, 722.56 &[10,8]\, 722.56 &[8,7]\, 747.38  &[30,36]\, 749.68   &[30,36]\, 749.67   & 780.77\\
AsN      & [4,4]1052.25  & [4,4]1085.11  & [4,4]1088.54  &[10,8]\, 934.37 &[10,8]\, 934.65  & [8,7]1055.11  &[40,41]1023.62    &[40,41]1023.89   &1068.54\\
As$_2$   & [4,4]\, 414.06  & [4,4]\, 413.52  & [4,4]\, 414.28  & [10,8]\, 388.45& [10,8]\, 388.42& [8,7]\, 410.69  &[66,54]\, 414.24  &[66,54]\, 414.24   & 430.00\\
Li$_2$   & [2,5]\, 388.38  & [2,5]\, 321.77  & [2,5]\, 315.30  & [2,5]\, 315.00& [6,10]\, 308.12& [2,5]\, 343.25  &[6,28]\, 337.73   &[6,28]\, 337.73   & 351.41\\
ClF      &[10,8]\, 752.08  &[10,7]\, 808.43  &[10,7]\, 689.23  &[14,10]\, 569.73& [14,8]\, 578.98&[12,9]\, 745.00  &[26,32]\, 717.57  &[26,32]\, 696.26   & 783.45\\
Cl$_2$   &[10,7]\, 539.06  &[10,7]\, 498.40  &[10,7]\, 504.52  &[14,10]\, 448.17& [14,8]\, 450.35&[12,9]\, 522.17 &[34,36]\, 509.86  &[34,36]\, 509.71   & 559.75\\
BrF      &[10,8]\, 649.39  &[10,7]\, 598.63  &[10,7]\, 625.71  &[14,10]\, 526.99& [14,8]\, 533.89&[12,9]\, 662.93  &[44,41]\, 619.66  &[44,41]\, 619.35   & 669.68\\
BrCl     & [6,5]\, 414.75  &[10,7]\, 410.42  &[10,7]\, 400.00  &[14,10]\, 424.17& [14,8]\, 366.81&[12,9]\, 422.58  &[52,45]\, 402.30  &[52,45]\, 402.07   & 444.32\\
Br$_2$   & [6,5]\, 306.19  & [8,6]\, 285.86  & [8,6]\, 288.04  & [12,9]\, 311.25& [14,8]\, 267.78&[12,9]\, 314.74  &[70,54]\, 294.13  &[70,54]\, 294.17   & 325.00\\
\hline%--------------------------------------------------------------------------------------------------------------------------------------\\
RMSD   &       60.55   &       54.07   &       41.44   &       123.68   &        98.11   &       46.83   & 43.74   & 41.73    &       \\        
MSD    &       11.18   &       -4.04   &       -9.75   &        -8.70   &       -72.73   &       -2.76   &-23.84   &-21.77    &       \\ 
MAD    &       36.16   &       37.99   &       28.53   &        82.27   &        80.88   &       30.25   & 34.60   & 32.10    &       \\
\end{tabular}
  \end{ruledtabular}
  \label{tab:Benchmark_HVF}
  \end{varwidth}} 
\end{table*}

\section{RESULTS}
\subsection{Performance of the Daubechies \pdfproof{w}avelet \pdfproof{b}asis \pdfproof{s}et} 
\label{subsec:EDQC}
We evaluate the performance of KS MOs generated from HF, PBE, and PBE0 XC functionals in the Daubechies wavelet basis set. To further reduce the qubit number requirement, we use a reduced active space based on a IEPA1 energy criterion in our proposed approach. 
In order to fully reveal the performance of the Daubechies wavelet basis set, we adopt the result of the EDQC method of the qubit Hamiltonian in the given active space to compare to the experimental value so that the error is fully attributed to the inadequacy of the basis set.

\subsubsection{H$_2$, LiH, \pdfproof{and} HF}
\begin{figure}[htbp]
\centering
\includegraphics[scale=0.6]{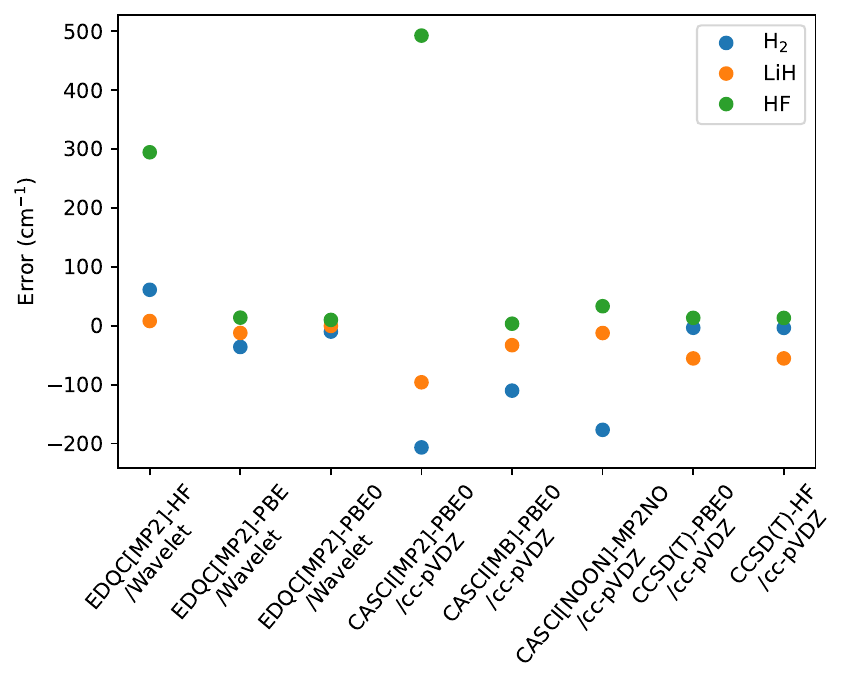}
\caption{Errors in the harmonic vibrational frequencies of H$_2$, LiH, and HF molecules (in cm$^{-1}$) calculated by a variety of methods. The errors (differences) are obtained by comparing the results to the corresponding experimental values.}
\label{fig:Error_HVF_H2LiHHF}
\end{figure}

To begin with, we choose three simple molecules, H$_2$, LiH and HF, from Table \ref{tab:Benchmark_HVF}, to benchmark the accuracies of the different methods and their respective errors in predicting the harmonic vibrational frequencies when compared to the experimental values (Fig.~\ref{fig:Error_HVF_H2LiHHF}).

For H$_2$, the approach EDQC[2,2]-XC/Wavelet (exact diagonalization method; the active space [2,2] determined by IEPA1) predicts vibrational frequencies of \pdfproof{4462.40} (61.19) cm$^{-1}$, \pdfproof{4365.33 ($-$35.88)} cm$^{-1}$, and \pdfproof{4391.45 ($-$9.76)} cm$^{-1}$ for XC = HF, PBE, and PBE0 functional, respectively, where the value inside the parenthesis denotes the error (difference) with respect to the experimental value and the same notation will be used in the following text. The improvement in the predicted vibrational frequency with the increasing level of the XC functionals is obvious.
%along with higher-level XC functionals. 
Notably, CASCI[2,2]-PBE0/cc-pVDZ (the active space [2,2] imitating the minimal basis set case), CASCI[2,2]-MP2NO/cc-pVDZ (using MP2 nature orbital; the active space [2,2] determined by NOON), and CASCI[2,3]-PBE0/cc-pVDZ (the active space [2,3] determined by IEPA1), yield vibrational frequencies of \pdfproof{4291.21 ($-$110.0)} cm$^{-1}$, \pdfproof{4224.68 ($-$176.53)} cm$^{-1}$, and \pdfproof{4194.85 ($-$206.35)} cm$^{-1}$, respectively, and all of which perform quite poorly. The results indicate that classical methods with a traditional atom-centered basis set do not perform well with a truncated small active space. The best performance in classical methods as expected is from CCSD(T)-HF/cc-pVDZ, which uses 10 MOs to yield a result of \pdfproof{4397.70 ($-$3.51)} cm$^{-1}$, while the result obtained by CCSD(T)-PBE0/cc-pVDZ is about the same.

For LiH, EDQC[2,3]-PBE0/Wavelet which gives \pdfproof{1405.36 ($-$0.14)} cm$^{-1}$ performs best, and furthermore the size of its active space is the smallest.
Similarly, for the HF (hydrogen fluoride) molecule, EDQC[2,3]-PBE/Wavelet which gives \pdfproof{4152.30} (13.91) cm$^{-1}$ and EDQC[2,3]-PBE0/Wavelet  which gives \pdfproof{4148.44} (10.04) cm$^{-1}$ both using only 3 MOs determined by IEPA1 outperform the other methods.

To sum up, for these three cases, the EDQC[IEPA1]-PBE0/Wavelet approach dominates the performance in both accuracy and efficiency at the same time.
%under the consideration of the accuracy and the efficiency at the same time. 
Clearly, the adaptation of a Daubechies wavelet basis set and a  small number of selected KS MOs can yield the vibrational frequencies as accurate as those obtained by the high-level WFT methods with a much larger number of MOs in the cc-pVDZ basis set. Therefore the KS Daubechies wavelet MOs in the active space selected by IEPA1 can provide significant improvement in describing molecular Hamiltonians in both size and quality. 

%For group 15 diatomic molecules (dimers composed of nitrogen family), EDQC[4,4]-XC/Wavelet with the active space composed of degenerate pi bondings and anti-bondings perform quite well (error deviations for all cases are smaller than 50 cm$^{-1}$) under the consideration of number of orbitals.

\subsubsection{Performance on the \pdfproof{o}verall \pdfproof{d}ataset}
\begin{figure}[htbp]
  \centering
  \includegraphics[width=8.6cm]{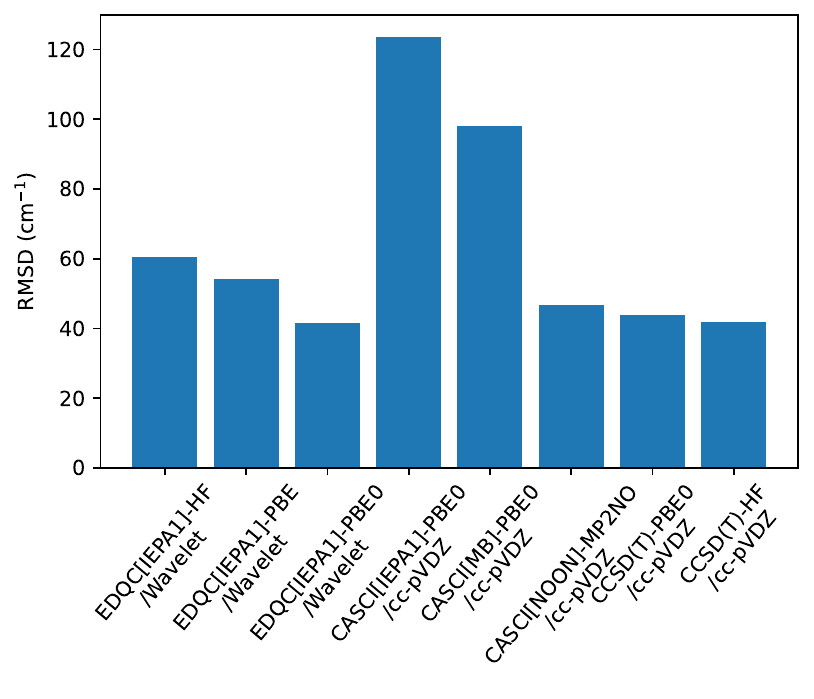}
  \caption{RMSD of the harmonic vibrational frequencies (in cm$^{-1}$) obtained by comparing the results to their corresponding experimental values for a variety of methods.}
  \label{fig:RMSD_HVF}
\end{figure}

\begin{figure*}[htbp!]
\centering
% add [] after \subfigure to show the label
\subfigure[\label{fig:SP_EDQC_MP2_HF}]{\includegraphics[scale=0.39]{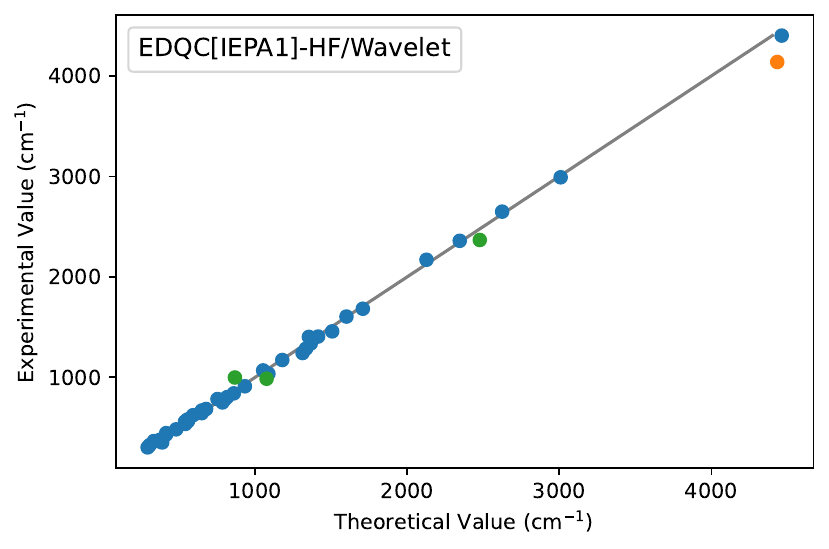}}\hfill
\subfigure[]{\includegraphics[scale=0.39]{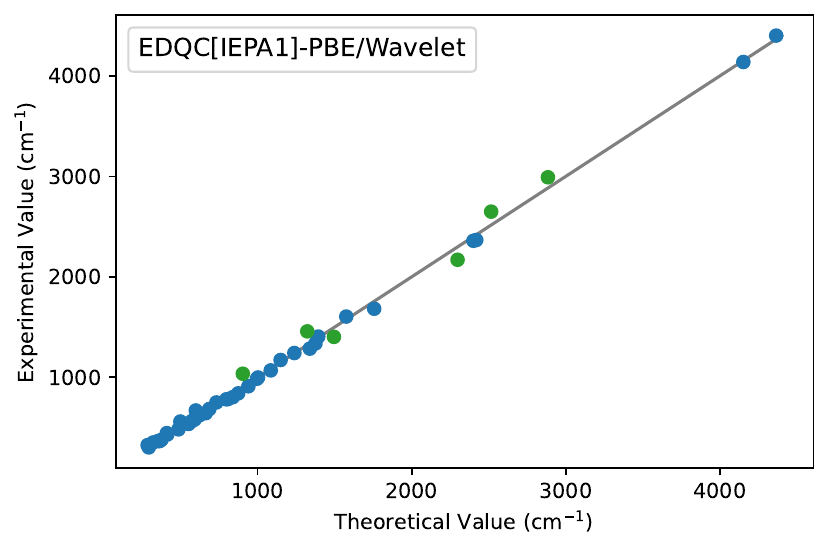}}\hfill
\subfigure[\label{fig:SP_EDQC_MP2_PBE0}]{\includegraphics[scale=0.39]{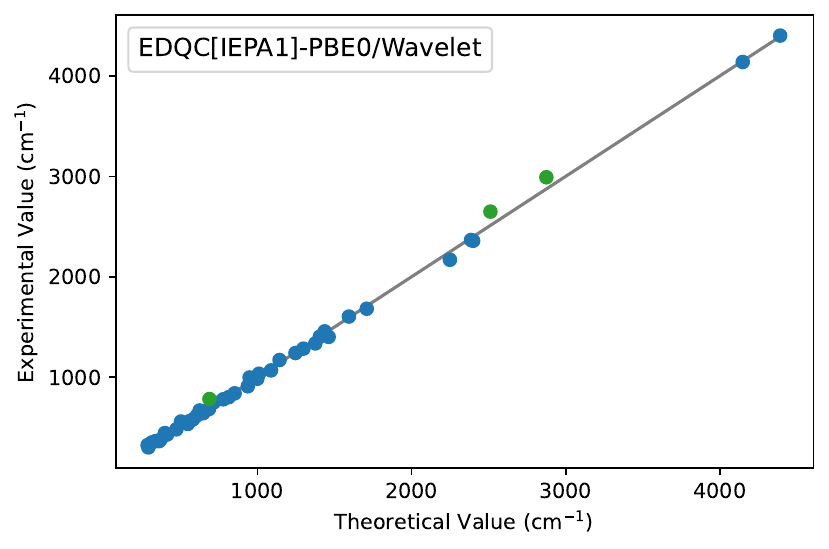}} \\
\subfigure[\label{fig:SP_CASCI_MP2_PBE0}]{\includegraphics[scale=0.39]{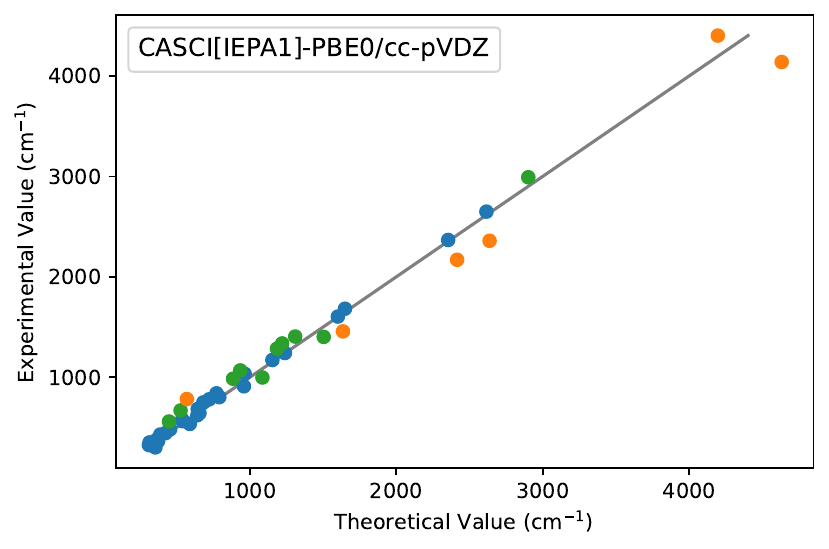}}\hfill
\subfigure[\label{fig:SP_CASCI_MB_PBE0}]{\includegraphics[scale=0.39]{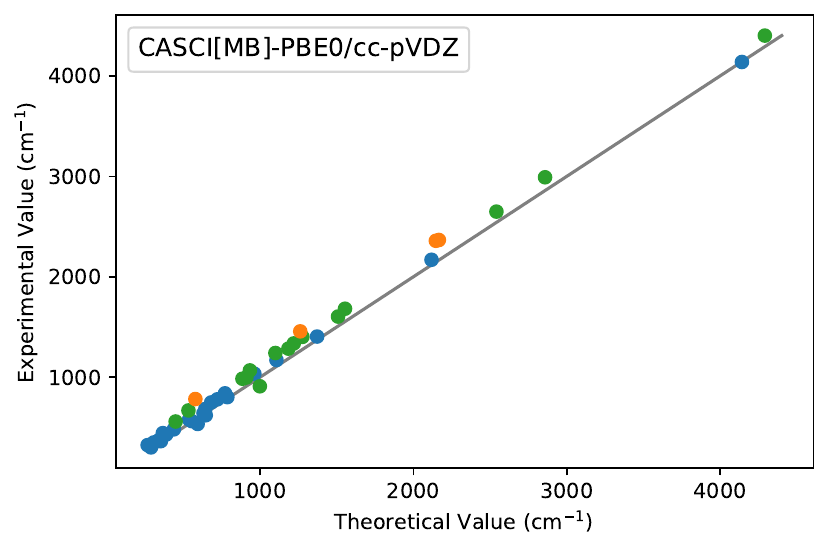}}\hfill
\subfigure[\label{fig:SP_CASCI_NO}]{\includegraphics[scale=0.39]{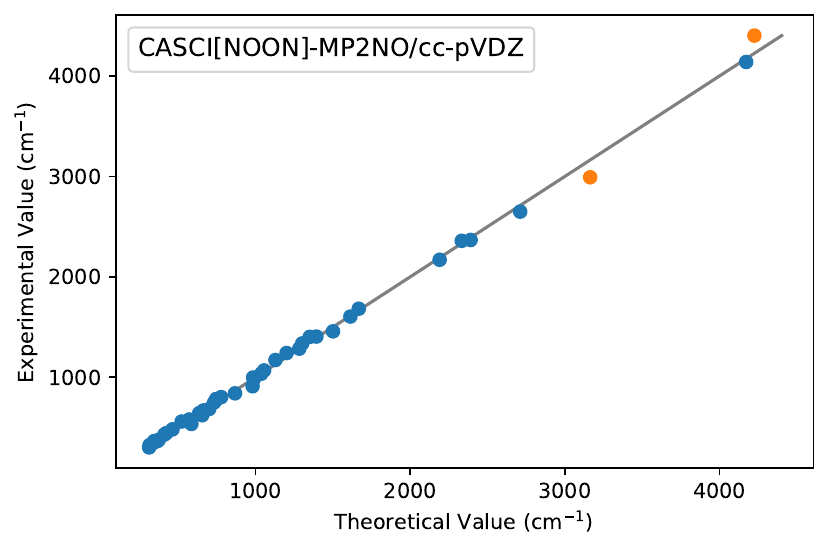}}\\
\subfigure[]{\includegraphics[scale=0.39]{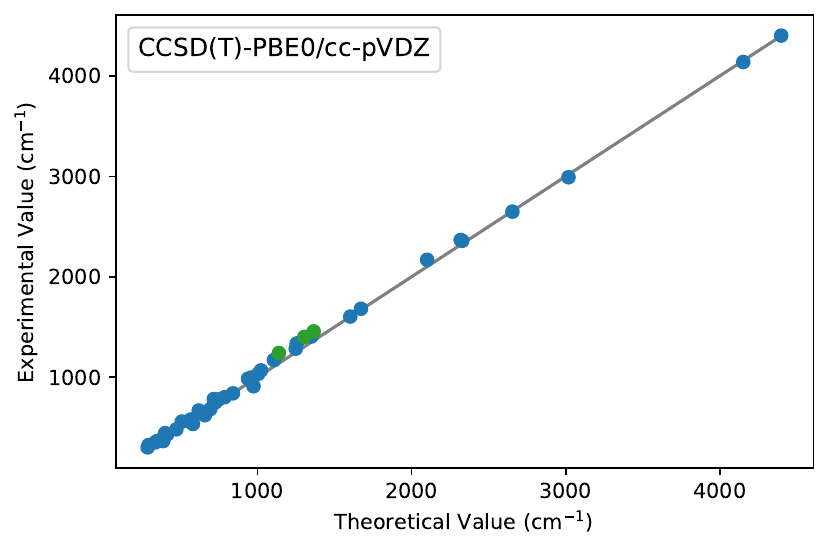}}
\subfigure[]{\includegraphics[scale=0.39]{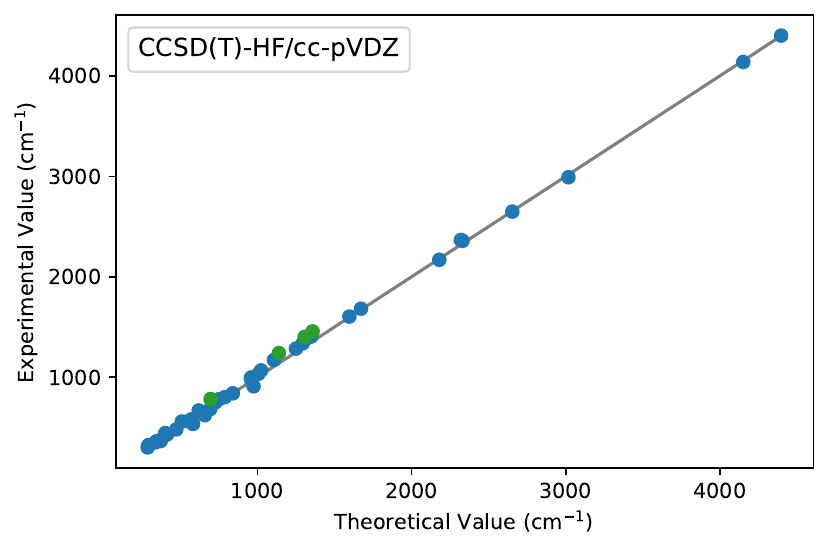}}
\caption{Scatter plots of the harmonic vibrational frequencies of the diatomic molecules versus their corresponding experimental values for each benchmark method. The blue, green, and orange dots denote outliers with absolute error deviation (in cm$^{-1}$) less than 85, between 85 and 150, and larger than 150, respectively.}
\label{fig:ScatterPlot_HVF}
\end{figure*}

To evaluate the overall performance of all the methods, we present statistical measures of errors of the harmonic vibrational frequencies including RMSD, %the root-mean-square deviation (RMSD), 
mean signed deviation (MSD), and mean absolute deviation (MAD) in Table \ref{tab:Benchmark_HVF}. In addition, the RMSD for each method is presented in Fig.~\ref{fig:RMSD_HVF}. Moreover, Fig.~\ref{fig:ScatterPlot_HVF} presents the scatter plots to display the correlation between theoretical predictions and experimental data in different approaches. 

As shown at the bottom of Table \ref{tab:Benchmark_HVF}, the EDQC[IEPA1]-XC/Wavelet approach gives the RMSD values of \pdfproof{60.55, 54.07}, and 41.44 cm$^{-1}$ for XC = HF, PBE, and PBE0 functional, respectively. Overall, the EDQC[IEPA1]-XC/Wavelet approach improves with the increasing level of the XC functionals.
%along with higher-level XC functionals. 
This trend is consistent with that of using the classical CCSD(T) method with MOs in similar XC functionals for the pruned closed-shell dataset in \pdfproof{Ref.~}\cite{Bertels:JCTC2021}.
As clearly seen from Table \ref{tab:Benchmark_HVF} and Fig.~\ref{fig:RMSD_HVF}, the approach with the best performance among all the methods is EDQC[IEPA1]-PBE0/Wavelet.
So EDQC[IEPA1]-PBE0/Wavelet is the approach proposed in this work, and it gives results in excellent agreement with the experimental data. 
We attribute its great performance to three factors:
(i) a better description of the Hamiltonian by introducing the Daubechies wavelets MOs, (ii)
%the further improvement of quality of MO and orbital orders by 
incorporating the electron correlation effect into the MOs via the XC functional,
(iii) a suitable selection of active space by IEPA1. In the following, we will discuss and emphasize these points through the comparison with relevant classical methods.
%The improvement by introducing the MOs expanded in the Daubechies wavelet basis set could be seen in the EDQC[\textcolor{magenta}{IEPA1}]-HF/Wavelet approach, where the RMSD is 41.06 cm$^{-1}$ if the outlier, the HF molecule (the orange point in Fig.~\ref{fig:SP_EDQC_MP2_HF}), is excluded.

The Daubechies wavelet MOs clearly outperform conventional cc-pVDZ basis set in our benchmark (see Fig.~\ref{fig:RMSD_HVF}).
For example, EDQC[IEPA1]-HF/Wavelet has a decent performance, and if the HF molecule [the orange point in Fig.~\ref{fig:SP_EDQC_MP2_HF}] is excluded, its RMSD value can be significantly improved to 41.06 cm$^{-1}$.
One of the errors for the outliers of EDQC[IEPA1]-HF/Wavelet 
might be due to the choice on the size of the initial truncated MO space, as more virtual orbitals generated by the HF XC functional outside the initial truncated MO space should be considered to reduce the error. On the other hand, if the electron correlation can be incorporated into the MOs, as in the EDQC[IEPA1]-XC/Wavelet method, using the same size of the initial truncated MO space could produce more accurate results than EDQC[IEPA1]-HF/Wavelet.
%is due to the restriction of the truncated virtual orbital space, where the order of \revise{MO} energies calculated from the HF XC functional \revise{for some molecules} would be incorrect such that some suitable \revise{MOs} would be outside the truncated virtual orbital space. 

For the KS MOs with XC functional beyond HF, the electron correlation effect is incorporated into the basis set via the XC functional. %\cite{Grimme:CPL1996,Grimme:JCP1999,Bour:CPL2001,Veseth:JCP2001,Beran:PCCP2003,Kim:PCCP2015,Kim:JCP2016,Bertels:JCTC2021}
As one can see from Fig.~\ref{fig:RMSD_HVF}, the RMSD values of EDQC[IEPA1]-XC/Wavelet can be reduced with the increasing level of the XC functionals from XC = HF to PBE and then to PBE0. 
The two outliers, HCl and HBr [the middle green points in Fig.~\ref{fig:SP_EDQC_MP2_PBE0}],
with errors larger than 100 cm$^{-1}$ for EDQC[IEPA1]-PBE0/Wavelet can be attributed to the over-corrections resulting in the error noncancellation because DFT-PBE0/Wavelet (presented in \pdfproof{Ref.~}\cite{SupplementalMaterial}) %; the equilibrium bond lengths are presented in Supplemental Material \cite{SupplementalMaterial}) 
and EDQC[IEPA1]-HF/Wavelet perform well for these two molecules.

%The EDQC[MB]-XC/Wavelet \revise{approaches whose active spaces imitate the minimal basis set do} not provide adequate results (see the data presented in Supplementary Table \ref{tab:HVF_EDQC_MB}) due to \modify{the consideration of the computational time cost for some large molecules that require a relatively large number of qubits (e.g., \textcolor{orange}{22} qubits) for simulations or due to} nonsmooth potential energy curves which could not give reliable harmonic vibrational frequencies. Compared with the EDQC[\textcolor{magenta}{IEPA1}]-XC/Wavelet approaches, all the EDQC[MB]-XC/Wavelet approaches have larger RMSD values, even \modify{excluding} those molecules with unavailable results. Thus, it is evident that selecting the active MOs correctly is essential in order to yield accurate results. After selecting a suitable active space, the number of MOs required to achieve the accurate results in the EDQC[\textcolor{magenta}{IEPA1}]-PBE0/Wavelet \modify{approach} is even less than the size of a minimal basis set, whereas the performance is significantly better.
Finally, it is evident that selecting the active MOs correctly is essential in order to yield accurate results with minimal computational cost. It is necessary to reduced the number of qubits required in order to make a calculation successful in the NISQ era, and we propose to utilize a IEPA1-based approach for active MO selection. The success of the EDQC[IEPA1]-XC/Wavelet approaches demonstrates the effectiveness of the IEPA1 MO selection method. In contrast, the EDQC[MB]-XC/Wavelet approaches whose active orbitals are selected in the order of ascending orbital energies do not provide adequate results (see the data presented in \pdfproof{Ref.~}\cite{SupplementalMaterial}). %; the equilibrium bond lengths are presented in Supplemental Material \cite{SupplementalMaterial}). 
Compared with the EDQC[IEPA1]-XC/Wavelet approaches, all the EDQC[MB]-XC/Wavelet approaches have significantly larger RMSD values, even though the number of active MOs used in these [MB] approaches are significantly higher.
 
To reveal the advantage of using the Daubechies wavelet basis set and IEPA1 active space selection, we compare the accuracy of the best achievable quantum computing method (EDQC[IEPA1]-PBE0/Wavelet) to that of the gold standard classical method (CCSD(T)-HF/cc-pVDZ, RMSD \pdfproof{=} 41.73 cm$^{-1}$).  The accuracies of these two approaches are comparable; however, the number of orbitals used in EDQC[IEPA1]-PBE0/Wavelet is much less than those in CCSD(T)-HF/cc-pVDZ, showing a polynomial advantage in the number of MOs. 
%Taking the N$_2$ family as an example, one can see that only 4 MOs are used in EDQC[IEPA1]-PBE0/Wavelet for quantum computation while the numbers of MOs required for CCSD(T)-HF/cc-pVDZ to achieve similar accuracies ranges from 28 to 54. In other word, in terms of the complexity in the number of terms in the Hamiltonian that need to be evaluated, EDQC[IEPA1]-PBE0/Wavelet in this case scales as $4^4=64$, while the scaling of CCSD(T) is $26^7 \approx 8.03\times 10^{9}$.
%%ranges from $28^7 \approx 1.35\times 10^{10}$ to $54^7\approx 1.34\times 10^{12}$.
\response{In the whole dataset, one can see that only 2 to 7 MOs are used in EDQC[IEPA1]-PBE0/Wavelet for quantum computing, whereas the CCSD(T)-HF/cc-pVDZ methods require 10 to 30 MOs in order to achieve similar accuracies. This difference in the number of orbitals leads to a significant change in computational complexity, as measured by the number of terms in the Hamiltonian that need to be evaluated. As a result, even in our scenario with a small number of MOs, the number of evaluations needed for the quantum computation of the EDQC[IEPA1]-PBE0/Wavelet approach scales as at most $(2\times7)^4=14^4 \approx 3.84 \times 10^4$ while the scaling of CCSD(T) is at most $30^7 \approx 2.19 \times 10^{10}$.}
This clearly demonstrates the potential of the quantum computing method in achieving a quantum advantage.

To display our proposed approach has its uniqueness, we compare EDQC[IEPA1]-PBE0/Wavelet to the classical corresponding method, CASCI[IEPA1]-PBE0/cc-pVDZ, where the dramatic difference in RMSD (the latter is 123.68 cm$^{-1}$) comes from the use of different basis sets, indicating that the same approach (IEPA1 active space and XC=PBE0) with the traditional basis sets is quite inaccurate. Moreover, we observe from Fig.~\ref{fig:SP_CASCI_MB_PBE0} that the results of the same classical approach with active space imitating the minimal basis set, CASCI[MB]-PBE0/cc-pVDZ, with RMSD being 98.11 cm$^{-1}$ are apparently red-shifted (see also the corresponding MSDs in Table \ref{tab:Benchmark_HVF}). 
On the other hand, both appreciably red-shifted and blue-shifted behaviors can be observed in Fig.~\ref{fig:SP_CASCI_MP2_PBE0} for quite a few molecules for the CASCI[EPA1]-PBE0/cc-pVDZ approach, showing that the active space determined by IEPA1 is not even useful for the traditional basis set.

The appropriate approach for classical CASCI method is to introduce the nature orbital. The accuracy of CASCI[NOON]-MP2NO/cc-pVDZ whose RMSD is 46.83 cm$^{-1}$ [the two obvious outliers are H$_2$ and HCl, marked by the orange points in Fig.~\ref{fig:SP_CASCI_NO}] is significantly better than that of CASCI[MB]-PBE0/cc-pVDZ or CASCI[IEPA1]-PBE0/cc-pVDZ and is comparable to that of EDQC[IEPA1]-PBE0/Wavelet. Note that the size of the active space selected by NOON is slightly larger than that by IEPA1.

The successful application of the KS MOs in the proposed EDQC[IEPA1]-PBE0/Wavelet approach might not be duplicated in the traditional basis set. The comparison between the two classical methods, CCSD(T)-PBE0/cc-pVDZ (RMSD \pdfproof{=} 43.74 cm$^{-1}$) and CCSD(T)-HF/cc-pVDZ (RMSD \pdfproof{=} 41.73 cm$^{-1}$) shows that the use of KS MOs with traditional basis set does not give much benefit in this closed-shell dataset (see also the result of $\Delta$(CCSD(T):RHF) in \cite{Bertels:JCTC2021}). Nevertheless, we remark that previous studies showed some kind of improvement for radicals \cite{Beran:PCCP2003} and open-shell systems \cite{Bertels:JCTC2021}.

\subsection{VQE(UCCSD) \pdfproof{b}enchmark}
\label{subsec:UCCSD}
For large molecular systems, the exact diagonalization method would not be feasible anymore. Instead, VQE is an algorithm that will be used practically on the near-term quantum computers. In Table \ref{tab:HVF_UCCSD-PBE0_Wavelet}, the harmonic vibrational frequencies obtained by VQE with the chemistry-inspired UCCSD ansatz and the SLSQP optimizer for the [IEPA1]-PBE0/Wavelet approach are presented (the equilibrium bond lengths are presented in \pdfproof{Ref.~}\cite{SupplementalMaterial}). The results show that the VQE(UCCSD) approach can be as accurate as the exact diagonalization method except for the BeO family, the CO family, and some of the N$_2$ family.

\begin{table}[h!]
\centering
\caption{Harmonic vibrational frequencies (in cm$^{-1}$) of neutral closed-shell diatomic molecules for VQE(UCCSD)[IEPA1]-PBE0/Wavelet. The approach EDQC[IEPA1]-PBE0/Wavelet is excerpted from Table \ref{tab:Benchmark_HVF} for the comparison. The value in the parenthesis denotes the difference in the results between VQE(UCCSD) and EDQC.}
\begin{ruledtabular}
\begin{tabular}{lcc}
      & EDQC[IEPA1]     & VQE(UCCSD)[IEPA1] \\  
Mol.  & -PBE0/Wavelet   & -PBE0/Wavelet     \\     
\hline%-----------------&---------------------
H$_2$ & [2,2]4391.45    & [2,2]4391.44   (\: $-$0.01) \\
LiH   & [2,3]1405.36    & [2,3]1405.34   (\: $-$0.02) \\
NaH   & [2,3]1143.76    & [2,3]1143.89   (\:\:\:\:  0.13) \\
BH    & [4,7]2386.10    & [4,7]2385.96   (\: $-$0.13) \\
AlH   & [4,7]1709.54    & [4,7]1712.38   (\:\:\:\:  3.05) \\
GaH   & [4,7]1594.36    & [4,7]1594.50   (\:\:\:\: 0.70) \\
HF    & [2,3]4148.44    & [2,3]4146.67   (\: $-$1.77) \\
HCl   & [2,3]2874.67    & [2,3]2874.69   (\:\:\:\:  0.03) \\
HBr   & [2,3]2511.84    & [2,3]2511.84   (\:\:\:\:  0.00) \\
LiF   & [4,6]\, 938.00  & [4,6]\, 937.94 (\: $-$0.06) \\
LiCl  & [4,4]\, 649.26  & [4,4]\, 649.26 (\:\:\:\:  0.00) \\
LiBr  & [6,5]\, 559.84  & [6,5]\, 559.81 (\: $-$0.03) \\
NaF   & [6,4]\, 548.67  & [6,4]\, 548.69 (\: $-$0.01) \\
NaCl  & [6,4]\, 362.88  & [6,4]\, 362.89 (\:\:\:\:  0.01) \\
NaBr  & [6,4]\, 292.99  & [6,4]\, 292.99 (\:\:\:\:  0.00) \\
BeO   & [6,6]1436.71    & [6,6]1508.87   (\:\: 72.16) \\
BeS   & [6,6]\, 949.54  & [6,6]\, 983.82 (\:\: 34.26) \\
BF    & [8,7]1462.33    & [8,7]1460.42   (\: $-$1.91) \\
BCl   & [8,6]\, 852.18  & [8,6]\, 849.61 (\: $-$2.57) \\
BBr   & [8,6]\, 686.03  & [8,6]\, 683.69 (\: $-$2.34) \\
AlF   & [8,7]\, 814.55  & [8,7]\, 814.80 (\:\:\:\: 0.25) \\
AlCl  & [4,4]\, 474.55  & [4,4]\, 474.55 (\:\:\:\: 0.00) \\
AlBr  & [4,4]\, 371.20  & [4,4]\, 371.20 (\:\:\:\: 0.00) \\
GaF   & [8,6]\, 609.27  & [8,6]\, 609.01 (\: $-$0.26) \\
GaCl  & [4,4]\, 340.81  & [4,4]\, 340.81 (\:\:\:\:  0.00) \\
CO    & [8,6]2248.86    & [8,6]2336.76   (\:\: 87.90) \\
CS    & [8,7]1298.67    & [8,7]1370.75   (\:\: 72.08) \\
CSe   & [8,7]1010.42    & [8,7]1056.19   (\:\: 45.77) \\
SiO   & [8,7]1247.22    & [8,7]1316.40   (\:\: 69.18) \\
SiS   & [8,8]\, 716.80  & [8,8]\, 747.32 (\:\: 30.52) \\
SiSe  & [8,6]\, 582.24  & [8,6]\, 613.98 (\:\: 31.75) \\
GeO   & [8,6]\, 999.61  & [8,6]1066.79   (\:\: 67.18) \\
N$_2$ & [4,4]2399.42    & [4,4]2402.52   (\:\:\:\:  3.10) \\
PN    & [4,4]1375.98    & [4,4]1372.77   (\: $-$3.21) \\
P$_2$ & [4,4]\, 778.53  & [4,4]\, 798.40 (\:\: 19.87) \\
AsN   & [4,4]1088.54    & [4,4]1055.19   ($-$33.35) \\
As$_2$& [4,4]\, 414.28  & [4,4]\, 414.10 (\: $-$0.18) \\
Li$_2$& [2,5]\, 315.30  & [2,5]\, 315.30 (\:\:\:\: 0.00)\, \\
ClF   &[10,7]\, 689.23\:\:  &[10,7]\, 689.11 (\: $-$0.12)\:\,\, \\
Cl$_2$&[10,7]\, 504.53\:\:  &[10,7]\, 504.56 (\:\:\:\: 0.04)\:\:\: \\
BrF   &[10,7]\, 625.71\:\:  &[10,7]\, 625.79 (\:\:\:\: 0.08)\:\:\: \\
BrCl  &[10,7]\, 400.00\:\:  &[10,7]\, 400.03 (\:\:\:\: 0.03)\:\:\: \\
Br$_2$& [8,6]\, 288.03  &[8,6]\, 288.16 (\:\:\:\: 0.12)\, \\ 
\hline%------&---------------&----------------------
RMSD  & \quad \quad \:\: 41.44  & 51.62  \quad \:  \\
MSD   &  \quad \quad \,\,\,  -9.73  & \, 1.70\quad \:\:\, \\
MAD   &  \quad \quad \,\, 28.52  & 35.37 \quad \, \\
\end{tabular}  
\end{ruledtabular}
\label{tab:HVF_UCCSD-PBE0_Wavelet} 
\end{table}

\begin{figure}[h]
  \centering
  \includegraphics[scale=0.59]{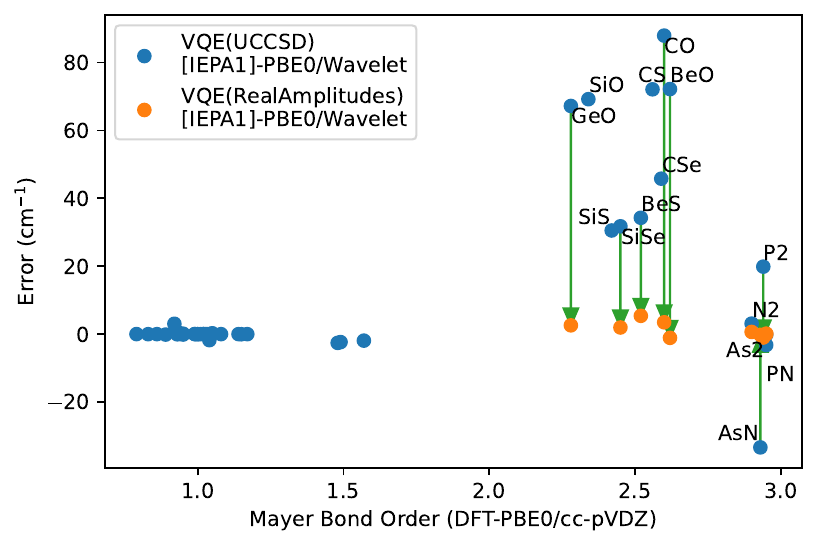}
  \caption{Mayer bond order indices calculated by DFT-PBE0/cc-pVDZ versus the error (difference) in the harmonic vibrational frequencies (in cm$^{-1}$) calculated by VQE(UCCSD)[IEPA1]-PBE0/Wavelet with respect to those by EDQC[IEPA1]-PBE0/Wavelet for the diatomic molecule in the benchmark dataset. The orange dots denote the vibrational frequency results for the specified molecules calculated by VQE(RealAmplitudes)[IEPA1]-PBE0/Wavelet. The relation between blue and orange dots is from Table \ref{tab:AnsatzComparison}, and the green arrows point toward the directions of improvement from the UCCSD ansatz to the RealAmplitudes ansatz.}
  \label{fig:ScatterPlot_MayerBondOrder}
\end{figure}

Previous study \cite{Lee:JCTC2019} showed that for systems with strongly correlated electrons, UCCSD would not give results achieving chemical accuracy even in the region near the equilibrium (bond-length) point. %even within the equilibrium region. 
In strongly correlated systems, the states resulting from the action of the UCCSD exponential operators that include only single and double coupled-cluster excitations might not encompass all those important configurations where the strongly correlated electrons would also be present.
This motivates us to investigate the Mayer bond order \cite{Mayer:CPL1983}, a good electron correlation descriptor applicable to multiconfigurational (strongly correlated) systems to quantify the degrees of bonding suitable for our analysis. That is, higher Mayer bond order indices correspond to more strongly correlated electrons. We calculate and show the Mayer bond order for the neutral closed-shell diatomic molecules in \pdfproof{Ref.~}\cite{SupplementalMaterial},
and the trend of the Mayer bond order is similar for different XC functionals considered here. We choose the Mayer \response{bond} order indices calculated by DFT-PBE0/cc-pVDZ (to be consistent with the XC functional and the traditional basis set used in this work) to present the relation with the harmonic vibrational frequencies calculated by VQE(UCCSD) in Fig.~\ref{fig:ScatterPlot_MayerBondOrder}. 
As clearly indicated in Fig.~\ref{fig:ScatterPlot_MayerBondOrder}, systems for which UCCSD does not yield accurate harmonic vibrational frequencies correspond to those whose Mayer bond order indices are larger than 2, which are the CO family, the N$_2$ family, and the BeO family (the blue points in the region where the Mayer bond order indices $>2$ in Fig.~\ref{fig:ScatterPlot_MayerBondOrder}), notably for the BeO family as they are traditionally thought of as the single-bond molecules. We conclude that the index of the Mayer bond order larger than 2 is a good descriptor to indicate that UCC truncated to SD might be too restricted to describe the harmonic vibrational frequency accurately.

\subsection{VQE(UCCSD) Versus VQE(RealAmplitudes)}
\label{subsec:AnsatzCf}

\begin{figure}[h]
  \hspace{-0.8cm}
  \includegraphics[scale=0.535]{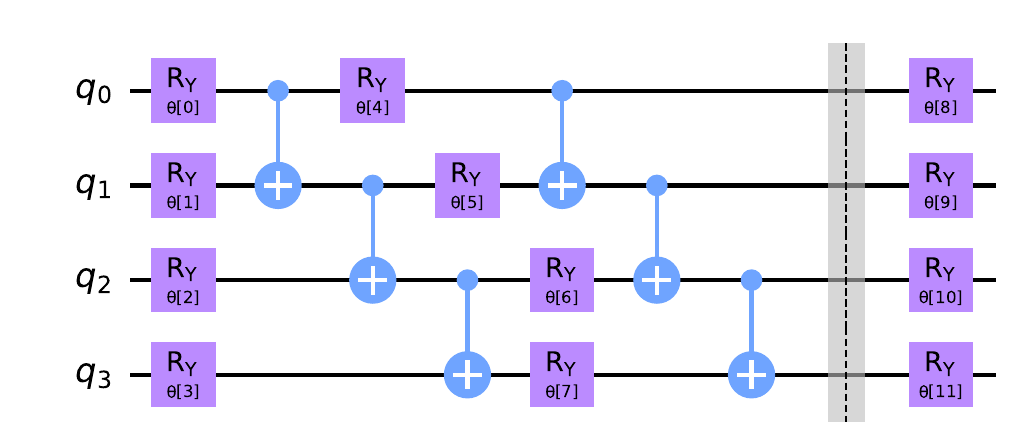}
 \caption{Typical quantum circuit of the hardware-efficient RealAmplitudes ansatz. For simplicity, a \pdfproof{four}-qubit two-local quantum circuit with \pdfproof{two} repeated unit pattern circuits is shown. The unit pattern circuit consists of a layer of parameterized R$_\text{Y}$ rotational gates applied on all qubits and a layer of CNOT gates in linear entanglement. For clarity, the barrier is added to separate the repeated unit pattern circuits from the final rotation layer and the Hartree-Fock initial state is skipped to draw. The rotational angles ${\theta}$s in the R$_\text{Y}$ gates denote the tunable circuit parameters.}
  \label{fig:RealAmp}
\end{figure}

\begin{table*}
  \centering
  \caption{\label{tab:AnsatzComparison} Comparisons of the harmonic vibrational frequencies and the relevant circuit information between the VQE(UCCSD) and VQE(RealAmplitudes) calculations using the Hamiltonian in the [IEPA1]-PBE0/Wavelet approach for the systems whose Mayer bond order indices are larger than 2 with the required number of qubits up to 10. The value inside the parenthesis in the harmonic vibrational frequency denotes the difference between VQE and EDQC.}
\begin{ruledtabular}
  \begin{threeparttable}[b]

%\begin{tabular}{lcccc||cccc}
\begin{tabular}{lcccccccc}
       &VQE(UCCSD)                   &      &   &State   &VQE(RealAmplitudes)   &Dep\tnote{a}    &    &State    \\  
Mol.   &[IEPA1]-PBE0/Wavelet           &Dep\tnote{a}  &N$\theta$\tnote{b} &Fidelity\tnote{c}  &[IEPA1]-PBE0/Wavelet  &(Rep\tnote{d}\ )  &N$\theta$\tnote{b} &Fidelity\tnote{c} \\
\hline
BeO    &[6,6]1508.87    (\:\: 72.15) &10914 &117 &0.98634 &[6,6]1435.63     ($-$1.09)  &99(30) &310 &0.9990823 \\
BeS    &[6,6]\, 983.82  (\:\: 34.27) &10914 &117 &0.99086 &[6,6]\, 955.98 (\:\: 5.39)  &99(30) &310 &0.9981542 \\
CO     &[8,6]2336.76    (\:\: 87.90) &8460  &92  &0.99729 &[8,6]2249.18   (\:\: 3.49)  &84(25) &260 &0.9999526 \\
SiSe   &[8,6]\, 613.98  (\:\: 31.75) &8460  &92  &0.99515 &[8,6]\, 584.19 (\:\: 1.96)  &99(30) &310 &0.9996153 \\
GeO    &[8,6]1066.79    (\:\: 67.18) &8460  &92  &0.99068 &[8,6]1012.25   (\:\: 2.59)  &84(25) &260 &0.9988251 \\
N$_2$  &[4,4]2402.52 (\:\:\:\: 3.10) &1480  &26  &0.99991 &[4,4]2400.07   (\:\: 0.65)  &35(10) & 66 &0.9999982 \\
PN     &[4,4]1372.77    (\: $-$3.21) &1480  &26  &0.99775 &[4,4]1376.12   (\:\: 0.14)  &35(10) & 66 &0.9999997 \\
P$_2$  &[4,4]\, 798.40  (\:\: 19.87) &1480  &26  &0.99988 &[4,4]\, 777.54   ($-$0.99)  &35(10) & 66 &0.9999974 \\
AsN    &[4,4]1055.19      ($-$33.35) &1480  &26  &0.99785 &[4,4]1088.34     ($-$0.20)  &35(10) & 66 &0.9999982 \\
As$_2$ &[4,4]\, 414.10  (\: $-$0.18) &1480  &26  &0.99971 &[4,4]\, 414.31 (\:\: 0.03)  &35(10) & 66 &0.9999996 \\
\end{tabular}

\begin{tablenotes}[flushleft]
  \item [a] Dep denotes the circuit depths.
  \item [b] N$\theta$ denotes the number of tunable circuit parameters.
  \item [c] State fidelity denotes the average state fidelity for the molecular distance points employed to calculate the vibrational frequency.
  \item [d] Rep denotes the number of repetitions of the unit pattern circuit.
\end{tablenotes}
\end{threeparttable}
  
\end{ruledtabular}
\end{table*}

For those systems whose Mayer bond order indices are larger than 2, we then consider a heuristic hardware-efficient ansatz, the RealAmplitudes ansatz (see Fig.~\ref{fig:RealAmp}) implemented in \pdfproof{\textsc{qiskit}} \cite{Qiskit}, since it can go beyond the restriction of the accessible Hilbert space of the chemistry-inspired UCCSD ansatz. 
That is, the hardware-efficient ansatz can increase its expressibility \cite{Sim:AQT2019} by increasing the number of repeated unit pattern circuit consisting of a layer of parameterized R$_\text{Y}$ rotational gates and a layer of entanglement circuit with two-qubit entangling gates (see Fig.~\ref{fig:RealAmp}). 
%The key feature of the hardware-efficient ansatz is that it can increase the expressibility by repeating the unit pattern \modify{circuit consisting of a layer of parameterized \textcolor{orange}{R$_\text{Y}$} rotational gates and a layer of linear entanglement circuit with two-local entangling gates}.
\response{During the procedure of increasing the number of repetitions, the results from the circuits with each number of repetitions will be inspected in order to evaluate and decide what the best circuit depth and number of tunable parameters for the molecule under consideration would be. Meanwhile, the performance of such non-chemistry-inspired ansatz is accomplished through the heuristic search on the parameter space.}

In Table \ref{tab:AnsatzComparison}, the results of the harmonic vibrational frequencies between the UCCSD ansatz and the RealAmplitudes ansatz with linear entanglement are compared for the systems whose Mayer bond order indices are larger than 2 and whose required numbers of qubits are not greater than 10 due to the computational time cost consideration.
%Furthermore, 
In order to accurately measure the degrees of errors 
%without being affected by the way of measurements 
and directly distinguish the accuracies achieved by the UCCSD ansatz and the RealAmplitudes ansatz in VQE, we calculate the state fidelity (defined as $|\langle \Phi_1 | \Phi_2 \rangle|^2$ for pure states $|\Phi_1\rangle$ and $|\Phi_2\rangle$) to measure how much overlap between the exact \pdfproof{wave function} obtained from the exact diagonalization method and the \pdfproof{wave function} obtained from the UCCSD ansatz and from the RealAmplitudes ansatz, respectively.
We use 10 repetitions of the unit pattern circuit for \pdfproof{six}-qubit molecular systems and 25 or 30 repetitions for \pdfproof{ten}-qubit systems in the RealAmplitudes ansatz to achieve the desired accuracy. As one of the superior advantage of the hardware-efficient ansatz, the circuit depth of the RealAmplitudes ansatz, which is 84 or 99 for \pdfproof{ten}-qubit systems, is significantly shallower than the circuit depth of the UCCSD ansatz, which is 8460 or 10914 for the corresponding \pdfproof{ten}-qubit systems. Despite having shallower circuit depths, the RealAmplitudes ansatz could still achieve higher state fidelities than the UCCSD ansatz, and for the cases with a small number of qubits outstanding performance can be achieved. Note that the energy differences between the result obtained by the exact diagonalization and that by the RealAmplitudes ansatz for the molecules are all within the chemical accuracy, which is not true when the UCCSD ansatz is used.
This is a clear indication that a heuristic hardware-efficient quantum circuit can span a state space larger than that spanned by the UCCSD method, a sign of polynomial quantum advantage. 
To the best of our knowledge, our investigation is the first systematical benchmark study to demonstrate that a heuristic hardware-efficient ansatz could outperform a chemistry-inspired UCCSD ansatz in predicting accurate molecular properties by quantum computation.

\response{At this moment, let us recapitulate the performance of quantum computing compared with gold standard method in classical computing. VQE(UCCSD)[IEPA1]-PBE0/Wavelet yields less accurate results compared with CCSD(T)-HF/cc-pVDZ (see Table \ref{tab:HVF_UCCSD-PBE0_Wavelet} and Table \ref{tab:Benchmark_HVF}), showing that using the UCCSD ansatz in quantum computation might not be preferable. However, VQE(RealAmplitudes)[IEPA1]-PBE0/Wavelet yields results basically very close or equivalent to the CCSD(T)-HF/cc-pVDZ method. This is especially notable for difficult-case molecules with Mayer bond order indices larger than 2 and whose required numbers of qubits are not greater than 10. In these cases, the quantum computing results could be as good as the best achievable accuracy obtained from the EDQC[IEPA1]-PBE0/Wavelet. Therefore quantum computation could achieve accuracy comparable with CCSD(T)-HF/cc-pVDZ, while, in terms of computational resources, the quantum algorithm benefited from our approach requires much smaller number of Hamiltonian evaluations compared with CCSD(T)-HF/cc-pVDZ.}

%As one of the superior advantage of the hardware-efficient ansatz, the circuit depth of the RealAmplitudes ansatz, which is 84 or 99 for \revise{10-qubit systems}, is significantly shallower than the circuit depth of the UCCSD ansatz, which is 8460 for \revise{the corresponding 10-qubit systems}. \st{under the same number of qubits.} 
We, however, note here that for the molecular systems using the same number of qubits, the number of tunable parameters for the RealAmplitudes ansatz is more than that for the UCCSD ansatz. For example, for the \pdfproof{ten}-qubit systems shown in Table \ref{tab:AnsatzComparison}, the number of tunable parameters for the RealAmplitudes ansatz is  260 or 310, while it is 92 or 117 for the UCCSD ansatz. Besides, 
the state fidelities of the RealAmplitudes ansatz become slightly lower when the number of qubits becomes larger. For a system using a large number of qubits for quantum computation, it is necessary to increase the number of repeated circuit layers of the parameterized and entanglement gates in order to obtain sufficient expressibility.
This could lead to a circuit with many parameters, making the optimization difficult as the initial values of the parameters would potentially critically affect the optimization result.
McClean \pdfproof{\textit{et al.}~}\cite{McClean:NatCommun2018} have shown that the optimization process of random initialization would be stuck in the local trap due to the barren plateaus.
Contrarily, the zero initialization (all the values of the tunable parameters are initialized to zeros) would give better results in most situations.
Specifically for the optimization procedure, we apply a combination of the SLSQP and L-BFGS-B optimizers, where the former converges faster but less accurate than the latter.
In the first stage, we consider the zero initialization of the tunable parameters and then also add small random numbers to them for the SLSQP optimizer to reach the converged parameters. Then, in the second stage, the converged parameters with relatively high state fidelities from SLSQP are taken as the initial parameters for L-BFGS-B to find the optimal parameter values. 

Since we aim to calculate the vibrational frequencies derived from the curvature around the equilibrium geometry of the PEC, we need to calculate the molecular ground state energies on a set of points (different distances between the atoms) around the equilibrium (bond-length) point  to construct the PEC. 
Note that in order to obtain accurate vibrational frequency for a molecule, an extremely high state fidelity might not be required to obtain a very accurate ground state energy; instated, the state fidelities or more precisely the ground state energies should have consistent correlated accuracy on all the points, which would give a parallel constant energy shift with respect to the reference PEC.
If achieving chemical accuracy is the only requirement, then the circuit could become even shallower to yield results within it. However, if the state fidelities are not high enough, one of the drawback of the non-chemistry-inspired ansatz is that each point is optimized independently so the optimized energy points on the PEC  behave like no appreciable correlation. To enhance the correlation between different molecular distance points, after the second optimization stage, if the optimization on one of the molecular distance points yields a distinctively lower molecular ground state energy, its converged parameters are taken as the initial parameters for the other points to optimize further until the variance of the energy for each point is small.
Therefore the decision on the choice of the number of the repeated circuit layers for the RealAmplitudes ansatz depends not only on the state fidelity of a single point but also on the variances of energies on all the points used to calculate the vibrational frequency.

We remark here that even for the same size of the RealAmplitudes circuit, different molecular systems (or different Hamiltonians) yield different degrees of state fidelities. This comes from the differences in structure and coefficients of the weighted Pauli terms in the Hamiltonians so that the optimizer may favor some cases. On the other hand, in addition to the number of repeated circuit layers, the types of the entangling gates or/and entanglement circuit structures would also affect the results. The advantage and disadvantage of more (or less) entanglement for different Hamiltonians remain to be clarified. The approach we present here is just a way to achieve high state fidelities, so potentially more efficient methods for large systems should be studied in the future.

%------------------------------------------------------------------------------
%  CONCLUSIONS AND OUTLOOK
%------------------------------------------------------------------------------
\section{CONCLUSION AND OUTLOOK} 
\label{sec:Conclusion}
We propose a quantum computational approach that combines KS MOs expanded in the Daubechies wavelet basis set and an optimal active space determined by IEPA1 energy criterion, resulting in a significantly reduced qubit number requirement (2 to 12 versus 20 to 60 required by cc-pVDZ with frozen core approximation) while maintaining excellent accuracy compared to the experimental data.
We validate the approach by benchmarking its performance on the harmonic vibrational frequencies of 43 neutral closed-shell diatomic molecules. 
The RMSD is small by using the Daubechies wavelets basis set and the error is further decreased by using KS MOs with higher-level XC functionals. The best approach  here is EDQC[IEPA1]-PBE0/Wavelet with performance comparable to CASCI[NOON]-MP2NO/cc-pVDZ and CCSD(T)-HF/cc-pVDZ.
The results obtained by this EDQC[IEPA1]-PBE0/Wavelet approach, considered as the best achievable results by quantum computation, are in great agreement with the experimental data. 
%In addition, its performance is comparable to CASCI[NOON]-MP2NO/cc-pVDZ and CCSD(T)-HF/cc-pVDZ, with significantly reduced qubit requirement. 
In contrast, for the traditional basis set on this closed-shell dataset, the same approach, e.g., CASCI[IEPA1]-PBE0/cc-pVDZ, could not provide noticeable improvements.

For larger systems, the exact diagonalization method is unfeasible due to the exponential scaling of the problem with the system size and thus a quantum computing approach is required.  
So we conduct VQE quantum computations of the vibrational frequencies of the 43  neutral closed-shell diatomic molecules using the Hamiltonians constructed from the approach of [IEPA1]-PBE0/Wavelet with the chemistry-inspired UCCSD ansatz. 
The results shows that the VQE(UCCSD) approach can be as accurate as the exact diagonalization method except for systems whose Mayer bond order indices are larger than 2. Then for those systems, we demonstrate that a hardware-efficient RealAmplitudes ansatz can provide significant improvements over the UCCSD ansatz.
This indicates that the hardware-efficient ansatz can avoid the restriction on the accessible Hilbert space by the chemistry-inspired UCCSD ansatz. At the same time, the appealing feature of the shallow circuit of the hardware-efficient ansatz will make quantum computation of accurate vibrational frequencies on the near-term NISQ devices realizable.
%\modify{To the best of our knowledge, our investigation is the first systematical benchmark study to demonstrate that a heuristic hardware-efficient ansatz could outperform a chemistry-inspired UCCSD ansatz in predicting accurate molecular properties by quantum computation.}

Based on the improvement of the results with the increasing hierarchy of the DFT XC functionals in this work, it is reasonable to expect that more promising results by quantum computing would be obtained by using higher-level XC functionals while keeping the number of qubits significant reduced. On the other hand, it is interesting to note that the XC functionals used here are developed for DFT and it might be worth trying to design XC functionals for KS MOs for the WFT calculation such that the accuracy and the efficiency of the calculation could be further improved reaching another level.

In summary, this benchmark study validates novel means to achieve highly-accurate calculations of molecular properties on quantum computers with significantly reduced qubit resources. Furthermore, VQE calculations with chemistry-inspired UCCSD and heuristic hardware-efficient ansatzes are compared to demonstrate the advantage of the heuristic ansatz in complex chemical bonding systems. Our calculations show that a quantum computer capable of carrying out calculations on $\leq$10 qubits with circuit depth $<$100 can accurately predict the vibrational frequencies of neutral closed-shell diatomic molecules, and these quantum resource requirements should be able to be achieved on near-term NISQ devices. In fact, according to the so-called $100\times100$ Challenge in the announcement of the IBM Quantum Summit 2022,  IBM Quantum plans to offer a tool able to calculate unbiased (noiseless) observables of circuits with 100 qubits and depth-100 gate operations in a reasonable runtime in 2024 \cite{IBMQ:100x100}. %(\url{https://research.ibm.com/blog/next-wave-quantum-centric-supercomputing}).
Our benchmark investigation here provides a critical assessment on the power of quantum computation of molecular properties and insights on further improvements.

%------------------------------------------------------------------------------
%  ACKNOWLEDGMENTS
%------------------------------------------------------------------------------
\section*{ACKNOWLEDGMENTS}
J.P.C. gratefully acknowledges the financial support from the National Science and Technology Council, Taiwan (NSTC 109-2112-M-018-008-MY3). 
Y.C.C. thanks the National Science and Technology Council, Taiwan (Grant No. NSTC 112-2119-M-002-018 and NSTC 111-2113-M-002-017), Physics Division, National Center for Theoretical Sciences (Grant No. 110-2124-M-002-012), and National Taiwan University (Grant No. 111L894603) for financial support. Y.C.C. is grateful to Computer and Information Networking Center, National Taiwan University for the support of high-performance computing facilities.
\pdfproof{A.H. gratefully acknowledges the sponsorship from Research Grants Council of the Hong Kong Special Administrative Region, China (Project No. CityU 11200120), City University of Hong Kong (Project No. 7005615, 7006103), CityU Seed Fund in Microelectronics (Project No. 9229135), and Hong Kong Institute for Advanced Study, City University of Hong Kong (Project No. 9360157).}
H.-S.G. acknowledges support from the National Science and
Technology Council, Taiwan under Grants 
No.~NSTC 112-2119-M-002-014, 
No.~NSTC 111-2119-M-002-006-MY3, 
No.~NSTC 111-2119-M-002-007,   
No.~NSTC 110-2627-M-002-002, 
No.~NSTC 111-2627-M-002-001,
and No.~NSTC 111-2627-M-002-006,
from the US Air Force Office of Scientific Research under
Award Number \pdfproof{FA2386-20-1-4052},
and from the National Taiwan University under
Grant No.~NTU-CC-112L893404. 
\pdfproof{H.-S.G. and Y.C.C. are} also grateful for the support from the
``Center for Advanced Computing and Imaging in Biomedicine (NTU-112L900702)'' through The Featured Areas Research Center Program within the framework of the Higher Education Sprout Project by the Ministry of Education (MOE), Taiwan, and
the support from the Physics Division, National
Center for Theoretical Sciences, Taiwan.

%------------------------------------------------------------------------------
%  Appendix
%------------------------------------------------------------------------------
%\appendix

%------------------------------------------------------------------------------
%  Reference
%------------------------------------------------------------------------------
% BibTex
%\bibliographystyle{}
\bibliography{Reference.bib} % Reference.bib

%------------------------------------------------------------------------------
%  Supplemental Material
%------------------------------------------------------------------------------

\onecolumngrid
\clearpage

\renewcommand{\thefigure}{S\arabic{figure}}
\renewcommand{\thetable}{S\arabic{table}}
\setcounter{page}{0}
\renewcommand{\thepage}{S\arabic{page}}
\setcounter{section}{0}
\renewcommand{\thesection}{S\Roman{section}}
%\renewcommand{\thesubsection}{S\Alph{subsection}}

% Title
\title{Supplemental Material: \\ \pdfproof{Accurate harmonic vibrational frequencies for diatomic molecules via \\ quantum computing}}

\maketitle
\onecolumngrid
%\tableofcontents

\vspace{1cm}
In this Supplementary Material, we provide additional data supplementary to those in the main text. Specifically, we first provide the results of the harmonic vibrational frequencies of the neutral closed-shell diatomic molecules calculated by the method of the exact diagonalization of the qubit Hamiltonian spanned in the minimal basis (MB) set of the Daubechies Wavelet molecule orbitals (MOs) obtained with different exchange (XC) functionals (denoted as EDQC[MB]-XC/Wavelet and presented in Table \ref{tab:HVF_EDQC_MB}), and calculated directly by the density functional theory (DFT) with different XC functionals (presented in Table \ref{tab:HVF_DFT-XC_Wavelet}). Then the equilibrium bond lengths of these molecules obtained by a variety of different methods considered in this work are presented in Sec.~\ref{sec:BL}. Finally, we present the results of the calculated Mayer order indices, serving as an electron correlation descriptors, for different XC functionals in Sec.~\ref{sec:MBO}.  

In all tables, as compared to the experimental data \cite{CCCBDB}, we present the statistical measures of errors of the harmonic vibrational frequencies or equilibrium bond lengths including the root-mean-square deviation (RMSD), mean signed deviation (MSD), and mean absolute deviation (MAD).

\clearpage
\section{Harmonic Vibrational Frequencies}
\subsection{EDQC[MB]-XC/Wavelet}
The active space imitates the MB set of the Daubechies wavelet MOs for different XC functionals would be the first attempt in the study of active space selection for quantum computation. However, this approach does not always converge toward the best solutions, and thus sometimes fails to produce smooth potential energy curves (PECs). For computation time consideration, molecules that requires qubit number over 22 are not calculated. Despite taking out those unavailable data, the RMSD of the calculated vibrational frequencies is still larger than for the [IEPA1] approach.

%The active space intimates the minimal basis set of the Daubechies Wavelet MOs for different XC functionals would be the first attempt in the study of active space selection for quantum computation. The data of the harmonic vibrational frequencies calculated using this approach denoted as EDQC[MB]-XC/Wavelet for the neutral closed-shell diatomic molecules investigated are presented in Supplementary Table \ref{tab:HVF_EDQC_MB}. However, the minimal basis set of the Daubechies Wavelet MOs does not always converge toward the best solutions and thus fails to produce smooth potential energy curves (PECs) for calculating the harmonic vibrational frequencies for some of the molecules out of the 43 neutral closed-shell diatomic molecules investigated. For computation time consideration, molecules that requires qubit number over 20 are not calculated. Despite taking out those unavailable data, the root-mean squared deviation (RMSD) of the calculated vibrational frequencies as compared to the experimental data for the remaining diatomic molecules is still larger than that for all the 43 diatomic molecules calculated by the MP2 active space selection method described in the main text. 

%We presented our proposed approach with this kind of active space in Supplemental Table \ref{tab:HVF_EDQC_MB} as the reference. This poor performance directly emphasizes on the successful MP2 selection method adopted in the main text.
% EDQC[MB]-XC/Wavelet: Harmonic Vibrational Frequencies
\begin{table}[htbp!]
  \centering
  \caption{Harmonic vibrational frequencies (in cm$^{-1}$) of the neutral closed-shell diatomic molecules calculated by the EDQC[MB]-XC/Wavelet method.}
  \renewcommand{\arraystretch}{0.92}% Tighter
  \begin{ruledtabular}
    \begin{threeparttable}[b]

\begin{tabular}{lrrrr}
        &      EDQC[MB]      &       EDQC[MB]    &       EDQC[MB]    & \\
Mol.    &   -HF/Wavelet      &   -PBE/Wavelet    &  -PBE0/Wavelet    &\quad   Expt.\\
\hline%-&------------------&-----------------&------------------&\quad ----------
H$_2$   &  [2,2]4462.40    &  [2,2]4365.33   &  [2,2]4391.45   &\quad 4401.21 \\
LiH     &  [4,6]1413.68    &  [4,6]1396.24   &  [4,6]1495.74   &\quad 1405.50 \\
NaH     & [10,9]1175.87    & [10,9]1447.98   & [10,9]1491.98   &\quad 1171.97 \\
BH      &  [4,5]2484.61    &  [4,5]2419.95   &  [4,5]2413.40   &\quad 2366.73 \\
AlH     &  [4,5]1707.98    &  [4,5]1679.30   &  [4,5]1713.01   &\quad 1682.38 \\
GaH     &  [4,5]1592.99    &  [4,5]1521.18   &  [4,5]1585.48   &\quad 1604.52 \\
HF      &  [8,5]4435.81    &  [8,5]4317.47   &  [8,5]4369.98   &\quad 4138.39 \\
HCl     &  [8,5]3124.05    &  [8,5]2830.52   &  [8,5]2773.21   &\quad 2990.93 \\
HBr     &  [8,5]2669.98    &  [8,5]2445.70   &  [8,5]2400.07   &\quad 2649.00 \\
LiF     & [10,9]\, 923.11  & [10,9]\, 957.72 & [10,9]\, 987.15        &\quad 910.57 \\
LiCl    & [10,9]\quad \, \,nan\tnote{a} & [10,9]\, 648.10   & [10,9]\, 652.25  &\quad 642.95 \\
LiBr    & [10,9]\, 562.18  & [10,9]\, 619.24 & [10,9]\quad \, \,nan   &\quad 563.00 \\
NaF     &[16,12]\qquad \, \, -\tnote{b}    &[16,12]\qquad \, \, -\tnote{b}   &[16,12]\qquad \, \, -\tnote{b}    &\quad 535.66\\
NaCl    &[16,12]\qquad \, \, -\tnote{b}    &[16,12]\qquad \, \, -\tnote{b}   &[16,12]\qquad \, \, -\tnote{b}    &\quad 364.68\\
NaBr    &[16,12]\qquad \, \, -\tnote{b}    &[16,12]\qquad \, \, -\tnote{b}   &[16,12]\qquad \, \, -\tnote{b}    &\quad 302.00\\
BeO     & [10,9]1578.58    & [10,9]1220.23   & [10,9]1363.96   &\quad 1457.09 \\
BeS     & [10,9]\, 932.55  & [10,9]\, 993.06 & [10,9]\, 931.09 &\quad  997.94 \\
BF      & [10,8]1419.05    & [10,8]1492.91   & [10,8]1462.11   &\quad 1402.16 \\
BCl     & [10,8]\, 856.38  & [10,8]\, 847.14 & [10,8]\, 880.60 &\quad 840.29 \\
BBr     & [10,8]\, 715.93  & [10,8]\, 643.46 & [10,8]\, 628.53 &\quad 684.31 \\
AlF     & [10,8]\, 824.07  & [10,8]\, 839.11 & [10,8]\, 814.67 &\quad 802.32 \\
AlCl    & [10,8]\, 467.42  & [10,8]\, 442.81 & [10,8]\, 380.02 &\quad 481.77 \\
AlBr    & [10,8]\, 342.19  & [10,8]\, 414.38 & [10,8]\, 339.08 &\quad 378.00 \\
GaF     & [10,8]\, 593.89  & [10,8]\, 666.73 & [10,8]\, 574.67 &\quad 622.20 \\
GaCl    & [10,8]\, 332.36  & [10,8]\, 430.83 & [10,8]\, 327.79 &\quad 365.30 \\
CO      & [10,8]2112.98    & [10,8]2300.72   & [10,8]2253.31   &\quad 2169.76 \\
CS      & [10,8]1322.27    & [10,8]1265.49   & [10,8]1311.61   &\quad 1285.16 \\
CSe     & [10,8]1091.60    & [10,8]1348.48   & [10,8]1041.09   &\quad 1035.36 \\
SiO     & [10,8]1314.29    & [10,8]1201.10   & [10,8]1248.11   &\quad 1241.54 \\
SiS     & [10,8]\, 776.59  & [10,8]\, 580.54 & [10,8]\, 686.28 &\quad 749.65 \\
SiSe    & [10,8]\, 557.79  & [10,8]\, 521.02 & [10,8]\, 565.61 &\quad 580.00 \\
GeO     & [10,8]1020.01    & [10,8]\, 876.00 & [10,8]\, 932.97 &\quad 985.50 \\
N$_2$   & [10,8]2368.11    & [10,8]2419.96   & [10,8]2420.98   &\quad 2358.57 \\
PN      & [10,8]1368.70    & [10,8]1296.85   & [10,8]1306.52   &\quad 1336.95 \\
P$_2$   & [10,8]\, 807.58  & [10,8]\, 801.64 & [10,8]\, 819.61 &\quad 780.77 \\
AsN     & [10,8]1034.44    & [10,8]\, 971.32 & [10,8]\quad \, \,nan &\quad 1068.54 \\
As$_2$  & [10,8]\, 428.54  & [10,8]\, 348.91 & [10,8]\, 343.78 &\quad 430.00 \\
Li$_2$  & [6,10]\, 284.48  & [6,10]\, 328.15 & [6,10]\, 320.83 &\quad 351.41 \\
ClF     & [14,8]\, 646.08  & [14,8]\, 645.82 & [14,8]\, 670.04 &\quad 783.45 \\
Cl$_2$  & [14,8]\, 497.62  & [14,8]\, 493.83 & [14,8]\, 495.45 &\quad 559.75 \\
BrF     & [14,8]\, 620.43  & [14,8]\, 595.44 & [14,8]\, 589.67 &\quad 669.68 \\
BrCl    & [14,8]\, 397.00  & [14,8]\, 391.55 & [14,8]\, 390.97 &\quad 444.32 \\
Br$_2$  & [14,8]\, 285.19  & [14,8]\, 279.71 & [14,8]\, 287.61 &\quad 325.00 \\
\hline%--------------------------------------&--------------------------
RMSD    &        71.90     &      110.20     &       99.65     & \\      
MSD     &        12.25     &      -10.20     &      -11.10     & \\
MAD     &        48.48     &       81.40     &       71.16     & \\
\end{tabular}

\begin{tablenotes}[flushleft]
  \item [a] nan denotes the unavailable data due to the non-smooth potential energy curve
  \item [b] - denotes the unavailable data due to time consuming
\end{tablenotes}
\end{threeparttable}
  \end{ruledtabular}
  \label{tab:HVF_EDQC_MB}
\end{table}

\begin{figure}[htbp!]
  \centering
  % add [] after \subfigure to show the label
  \subfigure{\includegraphics[scale=0.58]{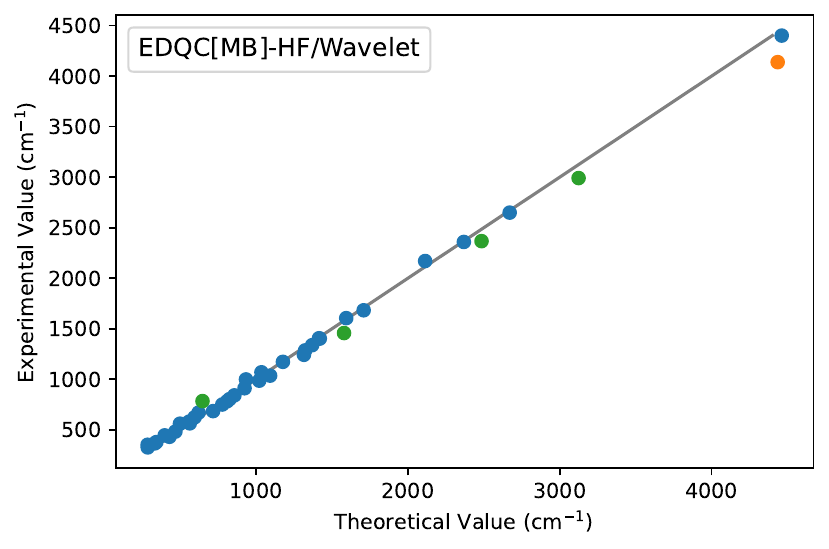}} \\
  \subfigure{\includegraphics[scale=0.58]{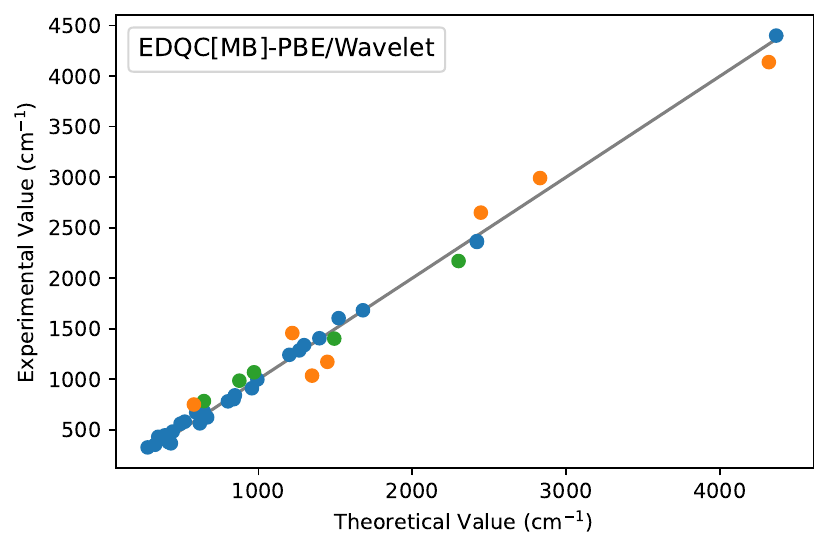}} \\
  \subfigure{\includegraphics[scale=0.58]{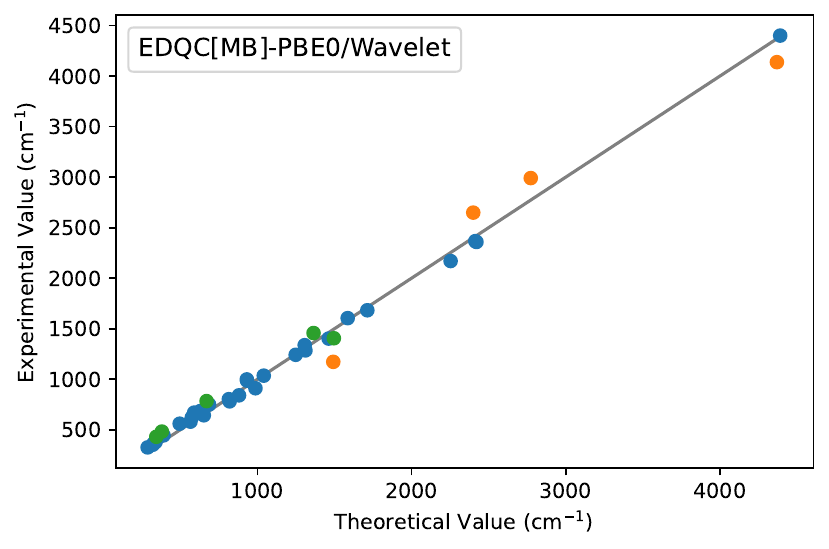}}
  \caption{The scatter plots of harmonic vibrational frequencies calculated by the EDQC[MB]-XC/Wavelet method. The blue, green, and orange dots denote outliers with absolute error deviation (in cm$^{-1}$) less than 85, between 85 and 150, and larger than 150, respectively.}
  \label{fig:ScatterPlot_HVF_EDQC_MB} % put after the \caption
\end{figure}

\clearpage
In Fig.~\ref{fig:ScatterPlot_HVF_EDQC_MB}, we present the scatter plots of EDQC[MB]-XC/Wavelet to visualize the deviation of the calculated vibrational frequencies from the experimental values for different XC functionals. 
%Different colors are used to emphasize that the amount of them directly affects the RMSD of the overall data.

\clearpage
\subsection{DFT-XC/Wavelet}
We present the results of vibrational frequencies obtained by the direct DFT calculations in Table \ref{tab:HVF_DFT-XC_Wavelet} as a reference to compare with those obtained by the WFT method using DFT molecular orbitals (MOs) where the electron correlation effect is incorporated into the basis set via the XC functional. 
%Then It is worth investigating such electron correlation effect would be over corrected or not in WFT. The DFT results provide the first factor that the chosen XC whether works.
% DFT-XC/Wavelet: Harmonic Vibrational Frequencies
\begin{table}[htbp!]
  \centering
  \caption{Harmonic vibrational frequencies (in cm$^{-1}$) of the neutral closed-shell diatomic molecules calculated by the DFT-XC/Wavelet method.}
  \begin{ruledtabular}
    \begin{tabular}{lrrrr}
       & DFT-HF   &DFT-PBE  &DFT-PBE0   &       \\
Mol.  &/Wavelet  &/Wavelet  &/Wavelet   &\qquad   Expt.\\
\hline%-------&----------&---------&-----\qquad ------&-------\\
H$_2$  &4637.47   &4316.68  & 4471.41   &\qquad 4401.21\\
LiH    &1440.89   &1387.45  & 1400.94   &\qquad 1405.50\\
NaH    &1185.95   &1149.07  & 1187.28   &\qquad 1171.97\\
BH     &2533.56   &2222.64  & 2369.50   &\qquad 2366.73\\
AlH    &1748.01   &1597.75  & 1641.09   &\qquad 1682.38\\
GaH    &1626.43   &1473.21  & 1541.16   &\qquad 1604.52\\
HF     &4442.78   &3887.81  & 4069.88   &\qquad 4138.39\\
HCl    &3142.79   &2867.42  & 2967.97   &\qquad 2990.93\\
HBr    &2808.47   &2567.17  & 2646.77   &\qquad 2649.00\\
LiF    & 931.91   & 961.69  &  809.52   &\qquad  910.57\\
LiCl   & 647.74   & 641.88  &  656.60   &\qquad  642.95\\
LiBr   & 557.99   & 549.93  &  562.89   &\qquad  563.00\\
NaF    & 545.59   & 606.79  &  478.26   &\qquad  535.66\\
NaCl   & 362.29   & 356.85  &  364.06   &\qquad  364.68\\
NaBr   & 292.55   & 291.28  &  297.62   &\qquad  302.00\\
BeO    &1720.70   &1446.31  & 1549.00   &\qquad 1457.09\\
BeS    &1065.85   & 979.73  & 1031.90   &\qquad  997.94\\
BF     &1520.84   &1346.83  & 1433.54   &\qquad 1402.16\\
BCl    & 882.18   & 806.15  &  850.73   &\qquad  840.29\\
BBr    & 690.98   & 669.10  &  686.25   &\qquad  684.31\\
AlF    & 839.18   & 670.81  &  802.97   &\qquad  802.32\\
AlCl   & 472.96   & 456.54  &  465.42   &\qquad  481.77\\
AlBr   & 371.72   & 363.10  &  368.78   &\qquad  378.00\\
GaF    & 605.42   & 610.59  &  595.70   &\qquad  622.20\\
GaCl   & 339.55   & 333.52  &  345.69   &\qquad  365.30\\
CO     &2427.79   &2100.73  & 2252.55   &\qquad 2169.76\\
CS     &1433.79   &1261.62  & 1347.17   &\qquad 1285.16\\
CSe    &1159.37   &1010.38  & 1067.33   &\qquad 1035.36\\
SiO    &1412.40   &1236.10  & 1235.68   &\qquad 1241.54\\
SiS    & 823.81   & 726.54  &  764.92   &\qquad  749.65\\
SiSe   & 636.02   & 552.48  &  580.82   &\qquad  580.00\\
GeO    &1092.04   & 892.66  &  958.38   &\qquad  985.50\\
N$_2$  &2677.82   &2355.13  & 2474.41   &\qquad 2358.57\\
PN     &1577.32   &1350.61  & 1442.19   &\qquad 1336.95\\
P2     & 927.11   & 771.37  &  832.11   &\qquad  780.77\\
AsN    &1277.04   &1059.44  & 1133.39   &\qquad 1068.54\\
As$_2$ & 500.03   & 426.07  &  413.52   &\qquad  430.00\\
Li$_2$ & 347.51   & 335.01  &  342.60   &\qquad  351.41\\
ClF    & 926.34   & 745.14  &  811.10   &\qquad  783.45\\
Cl$_2$ & 613.42   & 546.96  &  578.21   &\qquad  559.75\\
BrF    & 765.35   & 641.23  &  720.38   &\qquad  669.68\\
BrCl   & 490.42   & 432.87  &  455.42   &\qquad  444.32\\
Br$_2$ & 350.21   & 311.44  &  334.41   &\qquad  325.00\\
\hline%-------&----------&---------&-----------&-------
RMSD   & 131.37   &  66.37  &   45.82   & \\
MSD    &  91.52   & -37.21  &    9.84   & \\
MAD    &  95.16   &  43.54  &   32.93   & \\
\end{tabular}
  \end{ruledtabular}
  \label{tab:HVF_DFT-XC_Wavelet}
\end{table}

%%%%%%%%%%%%%%%%%%%%%%%%%%%%%%%%%%%%%%%%%%%%%%%%%%%%%%%%%%%%%
%%%%%%%%%%%%%%%%%%%%%%%%%%%%%%%%%%%%%%%%%%%%%%%%%%%%%%%%%%%%%
%%%%%%%%%%%%%%%%%%%%%%%%%%%%%%%%%%%%%%%%%%%%%%%%%%%%%%%%%%%%%

% Equilibrium Bond Lengths

\clearpage
\section{Equilibrium Bond Lengths}
\label{sec:BL}
%In the main text, we calculate the harmonic vibrational frequencies of molecules which should be evaluated at the corresponding equilibrium bond lengths. 
To accurately determine the harmonic vibrational frequency, it is important to know the equilibrium bond length of the molecule as it is calculated by the curvature of the PEC in the region near the equilibrium bond length point. Therefore, we present the equilibrium bond lengths in Table \ref{tab:Benchmark_BL} as complementary information of the harmonic vibrational frequencies shown in Table 1 in the main text. 
%since they are not main quantity to be compared.

The equilibrium bond lengths of the molecules calculated by the EDQC[MB]-XC/Wavelet, DFT-XC/Wavelet and  VQE(UCCSD)[IEPA1]-PBE0/Wavelet methods are shown in Table \ref{tab:BL_EDQC_MB}, Table \ref{tab:BL_DFT-XC_Wavelet}, and Table \ref{tab:BL_UCCSD-PBE0_Wavelet}, respectively.

\begin{table*}[htp]
  \centering
  \rotatebox{90}{
  \begin{varwidth}{\textheight}  
  \caption{Equilibrium bond lengths (in \text{\r{A}}) of the neutral closed-shell diatomic molecules obtained by different methods. \hfill} %\hfill to make caption left aligned
  \renewcommand{\arraystretch}{0.92}% Tighter
  \begin{ruledtabular}
    %\begin{turn}{90}
\begin{tabular}{lrrrrrrrrr}
         &        EDQC  &        EDQC  &        EDQC  &        CASCI  &        CASCI  &        CASCI  &  CCSD(T)       & CCSD(T)        &      \\
         &  [IEPA1]-HF  & [IEPA1]-PBE  &[IEPA1]-PBE0  & [IEPA1]-PBE0  &    [MB]-PBE0  & [NOON]-MP2NO  &   -PBE0        &    -HF         &      \\
Mol.     &    /Wavelet  &    /Wavelet  &    /Wavelet  &     /cc-pVDZ  &     /cc-pVDZ  &     /cc-pVDZ  &/cc-pVDZ        &/cc-pVDZ        & Expt.\\
\hline%--------------------------------------------------------------------------------------------------------------------------------------\\
H$_2$    & [2,2]0.7357  & [2,2]0.7434  & [2,2]0.7409  &  [2,3]0.7713  &  [2,2]0.7627  & [2,2]0.7693   & [2,10]0.7609   & [2,10]0.7609   &0.7414\\
LiH      & [2,3]1.6057  & [2,3]1.6074  & [2,3]1.6076  &  [2,5]1.6481  &  [4,6]1.6260  & [2,5]1.6140   & [4,19]1.6154   & [4,19]1.6154   &1.5949\\
NaH      & [2,3]1.9230  & [2,3]1.9309  & [2,3]1.9303  &  [2,5]1.9522  & [10,9]1.9416  & [2,5]1.9156   &[10,22]1.9217   &[10,22]1.9217   &1.8874\\
BH       & [4,5]1.2171  & [4,7]1.2289  & [4,7]1.2250  &  [4,9]1.2568  &  [4,5]1.2689  & [4,9]1.2501   & [4,18]1.2557   & [4,18]1.2558   &1.2324\\
AlH      & [4,6]1.6388  & [4,7]1.6469  & [4,7]1.6383  &  [4,9]1.6854  &  [4,5]1.6981  & [4,9]1.6565   & [4,18]1.6628   & [4,18]1.6623   &1.6478\\
GaH      & [4,6]1.6379  & [4,7]1.6380  & [4,7]1.6365  &  [4,8]1.7024  &  [4,5]1.7209  & [4,9]1.6796   & [4,18]1.6888   & [4,18]1.6860   &1.6630\\
HF       & [6,6]0.9001  & [2,3]0.9081  & [2,3]0.9088  & [8,10]0.8958  &  [8,5]0.9107  & [6,6]0.9231   & [8,18]0.9199   & [8,18]0.9199   &0.9168\\
HCl      & [2,3]1.2674  & [2,3]1.2790  & [2,3]1.2803  &  [8,9]1.2963  &  [8,5]1.3021  & [8,9]1.2737   & [8,18]1.2901   & [8,18]1.2899   &1.2746\\
HBr      & [2,4]1.4218  & [2,3]1.4348  & [2,3]1.4350  &  [8,9]1.4450  &  [8,5]1.4424  & [8,9]1.4160   & [8,18]1.4261   & [8,18]1.4258   &1.4144\\
LiF      & [4,6]1.5558  & [4,6]1.5553  & [4,6]1.5561  &  [8,7]1.5954  & [10,9]1.5774  & [8,7]1.5912   &[10,27]1.5892   &[10,27]1.5886   &1.5639\\
LiCl     & [4,6]2.0279  & [4,4]2.0223  & [4,4]2.0254  &  [8,7]2.0823  & [10,9]2.0793  & [6,6]2.0627   &[10,27]2.0823   &[10,27]2.0820   &2.0207\\
LiBr     & [6,5]2.1874  & [6,5]2.1819  & [6,5]2.1855  &  [8,7]2.2409  & [10,9]2.2256  & [6,8]2.2053   &[10,27]2.2318   &[10,27]2.2314   &2.1704\\
NaF      & [6,5]1.9307  & [6,4]1.9290  & [6,4]1.9298  &  [8,7]1.9104  &[16,12]1.8986  & [6,6]1.9130   &[16,30]1.9127   &[16,30]1.9120   &1.9259\\
NaCl     & [6,4]2.3869  & [6,4]2.3872  & [6,4]2.3875  &  [8,7]2.3896  &[16,12]2.3986  & [6,6]2.3919   &[16,30]2.4023   &[16,30]2.4021   &2.3608\\
NaBr     & [6,4]2.5433  & [6,4]2.5432  & [6,4]2.5440  &  [8,7]2.5127  &[16,12]2.5468  & [6,8]2.5126   &[16,30]2.5535   &[16,30]2.5532   &2.5020\\
BeO      & [6,9]1.3097  & [6,6]1.3605  & [6,6]1.3479  &  [8,6]1.3291  & [10,9]1.3899  & [6,6]1.3504   &[10,27]1.3673   &[10,27]1.3665   &1.3309\\
BeS      & [6,8]1.7828  & [6,6]1.7611  & [6,6]1.7666  &  [8,6]1.7425  & [10,9]1.7904  & [6,6]1.7684   &[10,27]1.7763   &[10,27]1.7761   &1.7415\\
BF       & [6,7]1.2679  & [8,7]1.2513  & [8,7]1.2569  & [10,8]1.2825  & [10,8]1.2866  & [8,7]1.2898   &[10,26]1.2960   &[10,26]1.2953   &1.2669\\
BCl      & [8,6]1.7121  & [8,6]1.7033  & [8,6]1.7054  & [10,8]1.7993  & [10,8]1.7993  & [8,7]1.7335   &[10,26]1.7468   &[10,26]1.7458   &1.7159\\
BBr      & [8,6]1.9029  & [8,6]1.8866  & [8,6]1.8889  & [10,8]1.9619  & [10,8]1.9619  & [8,7]1.9074   &[10,26]1.9211   &[10,26]1.9199   &1.8880\\
AlF      & [8,7]1.6276  & [8,7]1.6233  & [8,7]1.6257  & [10,8]1.7010  & [10,8]1.6984  & [8,7]1.7180   &[10,26]1.7112   &[10,26]1.7103   &1.6544\\
AlCl     & [2,3]2.1323  & [4,4]2.1216  & [4,4]2.1293  & [10,7]2.1956  & [10,8]2.2119  & [8,7]2.1889   &[10,26]2.1836   &[10,26]2.1826   &2.1301\\
AlBr     & [2,3]2.3049  & [4,4]2.3012  & [4,4]2.3096  & [10,7]2.3709  & [10,8]2.3865  & [8,7]2.3510   &[10,26]2.3521   &[10,26]2.3507   &2.2948\\
GaF      & [8,6]1.7065  & [8,6]1.6939  & [8,6]1.6995  & [10,8]1.7969  & [10,8]1.7947  & [8,7]1.8000   &[10,26]1.7909   &[10,26]1.7888   &1.7744\\
GaCl     & [4,4]2.2083  & [4,4]2.1930  & [4,4]2.2016  & [10,7]2.2602  & [10,8]2.2783  & [8,7]2.2428   &[10,26]2.2780   &[10,26]2.2494   &2.2017\\
CO       & [8,8]1.1350  & [8,6]1.1248  & [8,6]1.1278  & [10,8]1.1449  & [10,8]1.1374  & [8,7]1.1422   &[10,26]1.1450   &[10,26]1.1446   &1.1282\\
CS       & [8,6]1.5298  & [8,7]1.5244  & [8,7]1.5279  & [10,8]1.5754  & [10,8]1.5754  & [8,7]1.5606   &[10,26]1.5683   &[10,26]1.5671   &1.5349\\
CSe      &[10,7]1.6776  & [8,8]1.7009  & [8,7]1.6860  & [10,8]1.7225  & [10,8]1.7225  & [8,7]1.6992   &[10,26]1.7084   &[10,26]1.7064   &1.6762\\
SiO      & [8,8]1.4889  & [8,7]1.4944  & [8,7]1.4913  & [10,7]1.5349  & [10,8]1.5573  & [8,7]1.5512   &[10,26]1.5567   &[10,26]1.5555   &1.5097\\
SiS      & [8,8]1.9213  & [8,8]1.9390  & [8,8]1.9300  & [10,8]1.9880  & [10,8]1.9880  & [8,7]1.9717   &[10,26]1.9725   &[10,26]1.9715   &1.9293\\
SiSe     & [8,6]2.0531  & [8,6]2.0506  & [8,6]2.0517  & [10,8]2.1214  & [10,8]2.1214  & [8,7]2.1007   &[10,26]2.1045   &[10,26]2.1031   &2.0583\\
GeO      & [8,8]1.6014  & [8,6]1.5963  & [8,6]1.5957  & [10,8]1.6733  & [10,8]1.6733  & [8,7]1.6486   &[10,26]1.6697   &[10,26]1.6647   &1.6246\\
N$_2$    & [4,4]1.0986  & [4,4]1.0965  & [4,4]1.0981  & [10,8]1.1086  & [10,8]1.1241  & [8,7]1.1166   &[10,26]1.1192   &[10,26]1.1189   &1.0977\\
PN       & [4,4]1.4772  & [4,4]1.4795  & [4,4]1.4802  & [10,8]1.5319  & [10,8]1.5319  & [8,7]1.5229   &[10,26]1.5251   &[10,26]1.5245   &1.4909\\
P$_2$    & [4,4]1.8847  & [4,4]1.8877  & [4,4]1.8881  & [10,8]1.9525  & [10,8]1.9525  & [8,7]1.9394   &[10,26]1.9381   &[10,26]1.9373   &1.8934\\
AsN      & [4,4]1.6201  & [4,4]1.6135  & [4,4]1.6146  & [10,8]1.6670  & [10,8]1.6670  & [8,7]1.6443   &[10,26]1.6506   &[10,26]1.6489   &1.6184\\
As$_2$   & [4,4]2.1054  & [4,4]2.1115  & [4,4]2.1116  & [10,8]2.1652  & [10,8]2.1652  & [8,7]2.1458   &[10,26]2.1485   &[10,26]2.1465   &2.1026\\
Li$_2$   & [2,5]2.7930  & [2,5]2.6966  & [2,5]2.7269  &  [2,5]2.7650  & [6,10]2.7827  & [2,5]2.7277   & [6,28]2.7120   & [6,28]2.7120   &2.6730\\
ClF      &[10,8]1.6016  &[10,7]1.6611  &[10,7]1.6591  &[14,10]1.7587  & [14,8]1.7428  &[12,9]1.6777   &[14,26]1.6949   &[14,26]1.6951   &1.6283\\
Cl$_2$   &[10,7]2.0207  &[10,7]2.0312  &[10,7]2.0249  &[14,10]2.0403  & [14,8]2.0904  &[12,9]2.0360   &[14,26]2.0545   &[14,26]2.0543   &1.9879\\
BrF      &[10,8]1.7688  &[10,7]1.7871  &[10,7]1.7844  &[14,10]1.8657  & [14,8]1.8520  &[12,9]1.8002   &[14,26]1.8187   &[14,26]1.8187   &1.7589\\
BrCl     & [6,5]2.1821  &[10,7]2.1839  &[10,7]2.1821  &[14,10]2.1966  & [14,8]2.2335  &[12,9]2.1793   &[14,26]2.2010   &[14,26]2.2006   &2.1360\\
Br$_2$   & [6,5]2.3368  & [8,6]2.3438  & [8,6]2.3424  & [12,9]2.3444  & [14,8]2.3775  &[12,9]2.3215   &[14,26]2.3468   &[14,26]2.3462   &2.2811\\
\hline
RMSD     &      0.0291  &      0.0254  &      0.0252  &       0.0542  &       0.0605  &      0.0338   &       0.0424   &       0.0407   & \\      
MSD      &      0.0043  &      0.0049  &      0.0056  &       0.0450  &       0.0523  &      0.0294   &       0.0378   &       0.0364   & \\
MAD      &      0.0191  &      0.0185  &      0.0180  &       0.0468  &       0.0538  &      0.0301   &       0.0384   &       0.0370   & \\
\end{tabular}
%\end{turn}
 
  \end{ruledtabular}
  \label{tab:Benchmark_BL}
  \end{varwidth}} 
\end{table*}

% EDQC[MB]-XC/Wavelet: Equilibrium Bond Lengths
\clearpage
\subsection{EDQC[MB]-XC/Wavelet}
\begin{table}[h!]
  \centering
  \caption{Equilibrium bond lengths (in \text{\r{A}}) of the neutral closed-shell diatomic molecules calculated by the EDQC[MB]-XC/Wavelet method.}
  \begin{ruledtabular}
    \begin{threeparttable}[b]

\begin{tabular}{lrrrr}
        &     EDQC[MB]    &     EDQC[MB]   &     EDQC[MB]    & \\
Mol.    &  -HF/Wavelet    & -PBE/Wavelet   &-PBE0/Wavelet    &\quad  Expt.\\
\hline%-------------------------------------------------------\quad ---------
H$_2$   &  [2,2]0.7357    &  [2,2]0.7434   &  [2,2]0.7409    &\quad 0.7414\\
LiH     &  [4,6]1.6057    &  [4,6]1.6070   &  [4,6]1.6068    &\quad 1.5949\\
NaH     & [10,9]1.9231    & [10,9]1.9388   & [10,9]1.9309    &\quad 1.8874\\
BH      &  [4,5]1.2175    &  [4,5]1.2210   &  [4,5]1.2206    &\quad 1.2324\\
AlH     &  [4,5]1.6388    &  [4,5]1.6403   &  [4,5]1.6366    &\quad 1.6478\\
GaH     &  [4,5]1.6374    &  [4,5]1.6393   &  [4,5]1.6374    &\quad 1.6630\\
HF      &  [8,5]0.8999    &  [8,5]0.9020   &  [8,5]0.9010    &\quad 0.9168\\
HCl     &  [8,5]1.2598    &  [8,5]1.2829   &  [8,5]1.2760    &\quad 1.2746\\
HBr     &  [8,5]1.4069    &  [8,5]1.4401   &  [8,5]1.4373    &\quad 1.4144\\
LiF     & [10,9]1.5557    & [10,9]1.5542   & [10,9]1.5546    &\quad 1.5639\\
LiCl    & [10,9]\, \, \,nan\tnote{a}    & [10,9]2.0267   & [10,9]2.0272    &\quad 2.0207\\
LiBr    & [10,9]2.1989    & [10,9]2.1861   & [10,9]\, \, \,nan\tnote{a}    &\quad 2.1704\\
NaF     &[16,12]\qquad \, -\tnote{b}    &[16,12]\qquad \, -\tnote{b}   &[16,12]\qquad \, -\tnote{b}    &\quad 1.9259\\
NaCl    &[16,12]\qquad \, -\tnote{b}    &[16,12]\qquad \, -\tnote{b}   &[16,12]\qquad \, -\tnote{b}    &\quad 2.3608\\
NaBr    &[16,12]\qquad \, -\tnote{b}    &[16,12]\qquad \, -\tnote{b}   &[16,12]\qquad \, -\tnote{b}    &\quad 2.5020\\
BeO     & [10,9]1.3043    & [10,9]1.3660   & [10,9]1.3496    &\quad 1.3309\\
BeS     & [10,9]1.7507    & [10,9]1.7613   & [10,9]1.7671    &\quad 1.7415\\
BF      & [10,8]1.2571    & [10,8]1.2522   & [10,8]1.2574    &\quad 1.2669\\
BCl     & [10,8]1.7111    & [10,8]1.6995   & [10,8]1.7002    &\quad 1.7159\\
BBr     & [10,8]1.9023    & [10,8]2.0070   & [10,8]1.8901    &\quad 1.8880\\
AlF     & [10,8]1.6277    & [10,8]1.6235   & [10,8]1.6259    &\quad 1.6544\\
AlCl    & [10,8]2.1387    & [10,8]2.1515   & [10,8]2.1480    &\quad 2.1301\\
AlBr    & [10,8]2.3203    & [10,8]2.3424   & [10,8]2.3363    &\quad 2.2948\\
GaF     & [10,8]1.7070    & [10,8]1.7071   & [10,8]1.7171    &\quad 1.7744\\
GaCl    & [10,8]2.2093    & [10,8]2.2267   & [10,8]2.2245    &\quad 2.2017\\
CO      & [10,8]1.1309    & [10,8]1.1250   & [10,8]1.1281    &\quad 1.1282\\
CS      & [10,8]1.5316    & [10,8]1.5315   & [10,8]1.5260    &\quad 1.5349\\
CSe     & [10,8]1.6771    & [10,8]1.7680   & [10,8]1.6741    &\quad 1.6762\\
SiO     & [10,8]1.4770    & [10,8]1.4975   & [10,8]1.4930    &\quad 1.5097\\
SiS     & [10,8]1.9212    & [10,8]1.9562   & [10,8]1.9234    &\quad 1.9293\\
SiSe    & [10,8]2.0530    & [10,8]2.1068   & [10,8]2.1070    &\quad 2.0583\\
GeO     & [10,8]1.5895    & [10,8]1.6306   & [10,8]1.6188    &\quad 1.6246\\
N$_2$   & [10,8]1.1021    & [10,8]1.0967   & [10,8]1.0989    &\quad 1.0977\\
PN      & [10,8]1.4797    & [10,8]1.4936   & [10,8]1.4913    &\quad 1.4909\\
P$_2$   & [10,8]1.8859    & [10,8]1.8828   & [10,8]1.8844    &\quad 1.8934\\
AsN     & [10,8]1.6253    & [10,8]1.6575   & [10,8]\, \, \,nan\tnote{a}    &\quad 1.6184\\
As$_2$  & [10,8]2.1044    & [10,8]2.1637   & [10,8]2.1547    &\quad 2.1026\\
Li$_2$  & [6,10]2.7948    & [6,10]2.7086   & [6,10]2.7370    &\quad 2.6730\\
ClF     & [14,8]1.6393    & [14,8]1.6610   & [14,8]1.6588    &\quad 1.6283\\
Cl$_2$  & [14,8]2.0299    & [14,8]2.0287   & [14,8]2.0273    &\quad 1.9879\\
BrF     & [14,8]1.7704    & [14,8]1.7900   & [14,8]1.7872    &\quad 1.7589\\
BrCl    & [14,8]2.1860    & [14,8]2.1875   & [14,8]2.1861    &\quad 2.1360\\
Br$_2$  & [14,8]2.3447    & [14,8]2.3450   & [14,8]2.3446    &\quad 2.2811\\
\hline%-&---------------------------------------------------------------
RMSD    &       0.0308    &       0.0383   &       0.0287    & \\      
MSD     &       0.0030    &       0.0173   &       0.0095    & \\
MAD     &       0.0205    &       0.0287   &       0.0218    & \\
\end{tabular}

\begin{tablenotes}[flushleft]
  \item [a] nan denotes the unavailable data due to the non-smooth potential energy curve
  \item [b] - denotes the unavailable data due to time consuming
\end{tablenotes}
\end{threeparttable}
  \end{ruledtabular} 
  \label{tab:BL_EDQC_MB}
\end{table}

% DFT-XC/Wavelet: Equilibrium Bond Lengths
\clearpage
\subsection{DFT-XC/Wavelet}
\begin{table}[h!]
  \centering
  \caption{Equilibrium bond lengths (in \text{\r{A}}) of the neutral closed-shell diatomic molecules calculated by the DFT-XC/Wavelet method.}
 \begin{ruledtabular}
    \begin{tabular}{lrrrr}
       & DFT-HF  &DFT-PBE  &DFT-PBE0   & \\
Mol.  &/Wavelet &/Wavelet  &/Wavelet   &\qquad  Expt.\\
\hline%-------&---------&---------&-----\qquad ------&------\\
H$_2$  & 0.7338  & 0.7504  &  0.7448   &\qquad 0.7414\\
LiH    & 1.6044  & 1.6040  &  1.5965   &\qquad 1.5949\\
NaH    & 1.9187  & 1.8999  &  1.8935   &\qquad 1.8874\\
BH     & 1.2126  & 1.2451  &  1.2346   &\qquad 1.2324\\
AlH    & 1.6338  & 1.6747  &  1.6616   &\qquad 1.6478\\
GaH    & 1.6337  & 1.6767  &  1.6619   &\qquad 1.6630\\
HF     & 0.8996  & 0.9322  &  0.9192   &\qquad 0.9168\\
HCl    & 1.2587  & 1.2862  &  1.2756   &\qquad 1.2746\\
HBr    & 1.4034  & 1.4317  &  1.4205   &\qquad 1.4144\\
LiF    & 1.5556  & 1.5791  &  1.5624   &\qquad 1.5639\\
LiCl   & 2.0278  & 2.0215  &  2.0138   &\qquad 2.0207\\
LiBr   & 2.1875  & 2.1751  &  2.1682   &\qquad 2.1704\\
NaF    & 1.9306  & 1.9576  &  1.9417   &\qquad 1.9259\\
NaCl   & 2.3869  & 2.3759  &  2.3654   &\qquad 2.3608\\
NaBr   & 2.5433  & 2.5247  &  2.5146   &\qquad 2.5020\\
BeO    & 1.2900  & 1.3364  &  1.3157   &\qquad 1.3309\\
BeS    & 1.7167  & 1.7472  &  1.7315   &\qquad 1.7415\\
BF     & 1.2447  & 1.2748  &  1.2630   &\qquad 1.2669\\
BCl    & 1.7027  & 1.7262  &  1.7118   &\qquad 1.7159\\
BBr    & 1.8856  & 1.9039  &  1.8893   &\qquad 1.8880\\
AlF    & 1.6200  & 1.6680  &  1.6475   &\qquad 1.6544\\
AlCl   & 2.1258  & 2.1537  &  2.1379   &\qquad 2.1301\\
AlBr   & 2.3048  & 2.3239  &  2.3090   &\qquad 2.2948\\
GaF    & 1.6993  & 1.7526  &  1.7358   &\qquad 1.7744\\
GaCl   & 2.1984  & 2.2165  &  2.2015   &\qquad 2.2017\\
CO     & 1.1014  & 1.1355  &  1.1230   &\qquad 1.1282\\
CS     & 1.5007  & 1.5411  &  1.5244   &\qquad 1.5349\\
CSe    & 1.6426  & 1.6872  &  1.6681   &\qquad 1.6762\\
SiO    & 1.4623  & 1.5173  &  1.4972   &\qquad 1.5097\\
SiS    & 1.8918  & 1.9437  &  1.9227   &\qquad 1.9293\\
SiSe   & 2.0240  & 2.0775  &  2.0551   &\qquad 2.0583\\
GeO    & 1.5674  & 1.6317  &  1.6056   &\qquad 1.6246\\
N$_2$  & 1.0652  & 1.1018  &  1.0895   &\qquad 1.0977\\
PN     & 1.4358  & 1.4909  &  1.4715   &\qquad 1.4909\\
P2     & 1.8340  & 1.8983  &  1.8749   &\qquad 1.8934\\
AsN    & 1.5565  & 1.6195  &  1.5966   &\qquad 1.6184\\
As$_2$ & 2.0374  & 2.1096  &  2.1115   &\qquad 2.1026\\
Li$_2$ & 2.7809  & 2.7295  &  2.7284   &\qquad 2.6730\\
ClF    & 1.5747  & 1.6468  &  1.6125   &\qquad 1.6283\\
Cl$_2$ & 1.9570  & 1.9945  &  1.9692   &\qquad 1.9879\\
BrF    & 1.7169  & 1.7558  &  1.7569   &\qquad 1.7589\\
BrCl   & 2.1146  & 2.1532  &  2.1271   &\qquad 2.1360\\
Br$_2$ & 2.2680  & 2.3051  &  2.2788   &\qquad 2.2811\\
\hline%-------&---------&---------&-----------------
RMSD   & 0.0376  & 0.0169  &  0.0143   & \\ 
MSD    &-0.0185  & 0.0124  & -0.0027   & \\
MAD    & 0.0303  & 0.0135  &  0.0100   & \\
\end{tabular}  
 \end{ruledtabular}
 \label{tab:BL_DFT-XC_Wavelet}
\end{table}

% VQE(UCCSD): Equilibrium Bond Lengths
\clearpage
\subsection{VQE(UCCSD)}
\begin{table}[h!]
  \centering
  \caption{Equilibrium bond lengths (in \text{\r{A}}) of the neutral closed-shell diatomic molecules calculated by the VQE(UCCSD)[IEPA1]-PBE0/Wavelet method. The data of the EDQC approach are from Table \ref{tab:Benchmark_BL} for comparison. The value inside the parenthesis denotes the difference in bond lengths between the VQE(UCCSD) and EDQC methods.}
  \begin{ruledtabular} 
    \begin{tabular}{lrrr}
      &  EDQC[IEPA1]    & VQE(UCCSD)[IEPA1]           & \\
Mol.  & -PBE0/Wavelet   & -PBE0/Wavelet               &\qquad  Expt.\\
\hline
H$_2$ & [2,2]0.7409     & [2,2]0.7409 (\:\: 0.0000)   &\qquad 0.7414\\
LiH   & [2,3]1.6076     & [2,3]1.6076 (\:\: 0.0000)   &\qquad 1.5949\\
NaH   & [2,3]1.9303     & [2,3]1.9303 (\:\: 0.0000)   &\qquad 1.8874\\
BH    & [4,7]1.2250     & [4,7]1.2251 (\:\: 0.0001)   &\qquad 1.2324\\
AlH   & [4,7]1.6383     & [4,7]1.6381   ($-$0.0002)   &\qquad 1.6478\\
GaH   & [4,7]1.6365     & [4,7]1.6365 (\:\: 0.0000)   &\qquad 1.6630\\
HF    & [2,3]0.9088     & [2,3]0.9088 (\:\: 0.0000)   &\qquad 0.9168\\
HCl   & [2,3]1.2803     & [2,3]1.2803 (\:\: 0.0000)   &\qquad 1.2746\\
HBr   & [2,3]1.4350     & [2,3]1.4350 (\:\: 0.0000)   &\qquad 1.4144\\
LiF   & [4,6]1.5561     & [4,6]1.5560   ($-$0.0001)   &\qquad 1.5639\\
LiCl  & [4,4]2.0254     & [4,4]2.0255 (\:\: 0.0001)   &\qquad 2.0207\\
LiBr  & [6,5]2.1855     & [6,5]2.1856 (\:\: 0.0001)   &\qquad 2.1704\\
NaF   & [6,4]1.9298     & [6,4]1.9298 (\:\: 0.0000)   &\qquad 1.9259\\
NaCl  & [6,4]2.3875     & [6,4]2.3875 (\:\: 0.0000)   &\qquad 2.3608\\
NaBr  & [6,4]2.5440     & [6,4]2.5440 (\:\: 0.0000)   &\qquad 2.5020\\
BeO   & [6,6]1.3479     & [6,6]1.3181   ($-$0.0298)   &\qquad 1.3309\\
BeS   & [6,6]1.7666     & [6,6]1.7474   ($-$0.0192)   &\qquad 1.7415\\
BF    & [8,7]1.2569     & [8,7]1.2568   ($-$0.0001)   &\qquad 1.2669\\
BCl   & [8,6]1.7054     & [8,6]1.7069 (\:\: 0.0015)   &\qquad 1.7159\\
BBr   & [8,6]1.8889     & [8,6]1.8912 (\:\: 0.0023)   &\qquad 1.8880\\
AlF   & [8,7]1.6257     & [8,7]1.6256   ($-$0.0001)   &\qquad 1.6544\\
AlCl  & [4,4]2.1293     & [4,4]2.1293 (\:\: 0.0000)   &\qquad 2.1301\\
AlBr  & [4,4]2.3096     & [4,4]2.3096 (\:\: 0.0000)   &\qquad 2.2948\\
GaF   & [8,6]1.6995     & [8,6]1.6996 (\:\: 0.0001)   &\qquad 1.7744\\
GaCl  & [4,4]2.2016     & [4,4]2.2016 (\:\: 0.0000)   &\qquad 2.2017\\
CO    & [8,6]1.1278     & [8,6]1.1230   ($-$0.0048)   &\qquad 1.1282\\
CS    & [8,7]1.5279     & [8,7]1.5221   ($-$0.0058)   &\qquad 1.5349\\
CSe   & [8,7]1.6860     & [8,7]1.6763   ($-$0.0097)   &\qquad 1.6762\\
SiO   & [8,7]1.4913     & [8,7]1.4809   ($-$0.0104)   &\qquad 1.5097\\
SiS   & [8,8]1.9300     & [8,8]1.9194   ($-$0.0106)   &\qquad 1.9293\\
SiSe  & [8,6]2.0517     & [8,6]2.0419   ($-$0.0098)   &\qquad 2.0583\\
GeO   & [8,6]1.5957     & [8,6]1.5829   ($-$0.0128)   &\qquad 1.6246\\
N$_2$ & [4,4]1.0981     & [4,4]1.0977   ($-$0.0004)   &\qquad 1.0977\\
PN    & [4,4]1.4802     & [4,4]1.4807 (\:\: 0.0005)   &\qquad 1.4909\\
P$_2$ & [4,4]1.8882     & [4,4]1.8877   ($-$0.0005)   &\qquad 1.8934\\
AsN   & [4,4]1.6146     & [4,4]1.6153 (\:\: 0.0007)   &\qquad 1.6184\\
As$_2$& [4,4]2.1116     & [4,4]2.1109   ($-$0.0007)   &\qquad 2.1026\\
Li$_2$& [2,5]2.7269     & [2,5]2.7269 (\:\: 0.0000)   &\qquad 2.6730\\
ClF   &[10,7]1.6591     &[10,7]1.6591 (\:\: 0.0000)   &\qquad 1.6283\\
Cl$_2$&[10,7]2.0249     &[10,7]2.0249 (\:\: 0.0000)   &\qquad 1.9879\\
BrF   &[10,7]1.7844     &[10,7]1.7844 (\:\: 0.0000)   &\qquad 1.7589\\
BrCl  &[10,7]2.1821     &[10,7]2.1821 (\:\: 0.0000)   &\qquad 2.1360\\
Br$_2$& [8,6]2.3424     & [8,6]2.3424 (\:\: 0.0000)   &\qquad 2.2811\\
\hline
RMSD  &      0.0252     &      0.0257  \quad \quad \quad \quad \,\: & \\ 
MSD   &      0.0056     &      0.0030  \quad \quad \quad \quad \:\: & \\
MAD   &      0.0180     &      0.0184  \quad \quad \quad \quad \:\: & \\
\end{tabular} 
  \end{ruledtabular}
  \label{tab:BL_UCCSD-PBE0_Wavelet}
\end{table}

%%%%%%%%%%%%%%%%%%%%%%%%%%%%%%%%%%%%%%%%%%%%%%%%%%%%%%%%%%%%%
%%%%%%%%%%%%%%%%%%%%%%%%%%%%%%%%%%%%%%%%%%%%%%%%%%%%%%%%%%%%%
%%%%%%%%%%%%%%%%%%%%%%%%%%%%%%%%%%%%%%%%%%%%%%%%%%%%%%%%%%%%%

% Mayer Bond Order
\clearpage
\section{Mayer Bond Order}
\label{sec:MBO}
The Mayer bond order index between atoms A and B of a closed-shell molecule is defined as \cite{Mayer:CPL1983}:
\begin{equation}
  M_{AB} = \sum_{\mu \in A} \sum_{\nu \in B} (\mathbf{DS})_{\mu\nu} (\mathbf{DS})_{\nu\mu},   
\end{equation}
where $\mu,\nu$ are indices for the basis functions belonged to the assigned atom, %$\mathbf{D}$ is the spinless density matrix and $\mathbf{S}$ is the overlap matrix.
and $\mathbf{DS}$ denotes the product the spinless density matrix $\mathbf{D}$ and the overlap matrix $\mathbf{S}$.

In Table \ref{tab:MayerBondOrder}, we present the calculated results of the Mayer bond order for different XC functionals. The Mayer bond order can serve as a descriptor to characterize the electron correlations of the molecules. Note that the trend of the Mayer bond order is similar for different XC functionals.

\begin{table}[htbp!]
  \centering
  \caption{The Mayer bond order indices of neutral closed-shell diatomic molecules for DFT-XC/cc-pVDZ. \hfill} %\fhill to make caption left aligned
  \begin{ruledtabular}
    \begin{tabular}{lccc}
        &DFT-HF       &DFT-PBE    &DFT-PBE0 \\
Mol.    &/cc-pVDZ     &/cc-pVDZ   &/cc-pVDZ \\
\hline
H$_2$   & 1.00        & 1.00      & 1.00    \\
LiH     & 0.99        & 1.00      & 1.00    \\
NaH     & 0.92        & 0.96      & 0.95    \\
BH      & 0.91        & 0.89      & 0.89    \\
AlH     & 0.90        & 0.93      & 0.92    \\
GaH     & 0.91        & 0.94      & 0.93    \\
HF      & 1.03        & 1.05      & 1.04    \\
HCl     & 0.99        & 0.99      & 0.99    \\
HBr     & 0.99        & 0.99      & 0.99    \\
LiF     & 0.71        & 1.04      & 0.93    \\
LiCl    & 0.93        & 1.11      & 1.04    \\
LiBr    & 1.03        & 1.21      & 1.15    \\
NaF     & 0.59        & 0.97      & 0.83    \\
NaCl    & 0.66        & 0.88      & 0.79    \\
NaBr    & 0.74        & 0.95      & 0.86    \\
BeO     & 2.47        & 2.69      & 2.62    \\
BeS     & 2.44        & 2.56      & 2.52    \\
BF      & 1.39        & 1.63      & 1.57    \\
BCl     & 1.35        & 1.53      & 1.48    \\
BBr     & 1.33        & 1.54      & 1.49    \\
AlF     & 0.81        & 1.16      & 1.05    \\
AlCl    & 0.99        & 1.20      & 1.14    \\
AlBr    & 1.01        & 1.23      & 1.17    \\
GaF     & 0.78        & 1.14      & 1.03    \\
GaCl    & 0.91        & 1.14      & 1.08    \\
CO      & 2.52        & 2.63      & 2.60    \\
CS      & 2.52        & 2.57      & 2.56    \\
CSe     & 2.51        & 2.61      & 2.59    \\
SiO     & 2.18        & 2.40      & 2.34    \\
SiS     & 2.33        & 2.45      & 2.42    \\
SiSe    & 2.36        & 2.48      & 2.45    \\
GeO     & 2.08        & 2.34      & 2.28    \\
N$_2$   & 2.93        & 2.90      & 2.90    \\
PN      & 2.94        & 2.95      & 2.95    \\
P$_2$   & 2.98        & 2.93      & 2.94    \\
AsN     & 2.91        & 2.94      & 2.93    \\
As$_2$  & 2.98        & 2.95      & 2.95    \\
Li$_2$  & 1.01        & 1.01      & 1.01    \\
ClF     & 0.87        & 0.98      & 0.95    \\
Cl$_2$  & 0.98        & 1.03      & 1.02    \\
BrF     & 0.84        & 0.98      & 0.94    \\
BrCl    & 0.99        & 1.04      & 1.03    \\
Br$_2$  & 1.02        & 1.06      & 1.05    \\
\end{tabular}  
  \end{ruledtabular}
  \label{tab:MayerBondOrder}
\end{table}

%%%%%%%%%%%%%%%%%%%%%%%%%%%%%%%%%%%%%%%%%%%%%%%%%%%%%%%%%%%%%
%%%%%%%%%%%%%%%%%%%%%%%%%%%%%%%%%%%%%%%%%%%%%%%%%%%%%%%%%%%%%
%%%%%%%%%%%%%%%%%%%%%%%%%%%%%%%%%%%%%%%%%%%%%%%%%%%%%%%%%%%%%

% Reference_SM
\clearpage
%\bibliography{Reference_SM.bib}
%\input{SupplementalMaterial.bbl} % show the references only cited in Supplemental Material

% Paste the content of SupplementalMaterial.bbl here.
% Use {unsrt} to reorder the references.
%

\end{document}